\documentclass[useAMS,usenatbib]{mn2e}
\usepackage{epsfig}

\title{Molecular  cooling in the diffuse interstellar medium}
\author[Glover \& Clark]
{Simon C. O. Glover\thanks{E-mail: glover@uni-heidelberg.de} \& Paul C. Clark \\
\\ Universit\"at Heidelberg, Zentrum f\"ur Astronomie, Institut f\"ur Theoretische Astrophysik, \\ Albert-Ueberle-Strasse 2,  69120 Heidelberg, Germany
}

\begin{document}
\maketitle

\begin{abstract}
We use a simple one-zone model of the thermal and chemical evolution of interstellar gas to
study whether molecular hydrogen (H$_2$) is ever an important coolant of the warm,
diffuse interstellar medium (ISM). We demonstrate that at solar metallicity, H$_{2}$ cooling
is unimportant and the thermal evolution of the ISM is dominated by metal line cooling. At
metallicities  below $0.1 \: {\rm Z_{\odot}}$, however, metal line cooling of low density gas
quickly becomes unimportant and H$_2$ can become the dominant coolant, even though
its abundance in the gas remains small. We investigate the conditions required in order for
H$_2$ to dominate, and show that it provides significant cooling only when the ratio of
the interstellar radiation field strength to the gas density is small. Finally, we demonstrate
that our results are insensitive to changes in the initial fractional ionization of the gas
or to uncertainties in the nature of the dust present in the low-metallicity ISM.
\end{abstract}


\section{Introduction}
Observations of star formation within the Milky Way and other nearby metal-rich galaxies show that 
the surface density of star formation correlates better with the surface density of molecular gas than
with the total surface density or the surface density of atomic gas (\citealt{leroy08,bigiel08,bigiel11,schruba11};
although see also \citealt{shetty13}).
It is natural to assume that this correlation arises because the presence of molecular gas is a necessary 
prerequisite for star formation, but recent studies by \citet{klm11}, \citet{krum12} and \citet{gc12a,gc12b}
have shown that this interpretation is incorrect. Numerical simulations of star formation in dense
gas clouds have been used to demonstrate that the star formation rate in these clouds is largely insensitive
to the composition of the gas. If molecule formation and molecular cooling are artificially suppressed,
the predicted star formation rate that one obtains is very similar to that coming from models that include
molecular cooling \citep{gc12a}. These models suggest that the 
observed correlation between molecular gas and star formation comes about because the conditions that 
favour star formation also favour the formation of molecules. Furthermore, they predict that this correlation will
break down at low metallicities \citep{gc12b,krum12}.

However, these studies focussed on the thermal physics of individual gas clouds that were
assumed to have already been assembled from the diffuse interstellar medium (ISM). They
did not address the issue of whether molecular gas can play an important role at an earlier
stage of the process, prior to, or during the assembly of, these dense, cold clouds.

This question has been examined to some extent by a few previous numerical studies.
\citet{jappsen07} performed simulations of protogalaxy formation using a simplified
chemical model that accounted for H$_{2}$ formation and destruction, as well as the
ionization and recombination of C, O and Si, and showed that under these conditions, 
H$_{2}$ cooling dominates over metal line cooling at low densities ($n \sim 1 \: {\rm cm^{-3}}$ 
and below) for metallicities ${\rm Z} < 0.1 \: {\rm Z_{\odot}}$. However, their study did not include 
the effects of a pre-existing background radiation field, which is arguably a reasonable 
approach when considering the formation of the earliest protogalaxies, but which is 
unlikely to be valid for low metallicity galaxies in the local Universe.

More recently, \citet{gk11} reported that in their galaxy formation simulations, H$_{2}$ cooling
dominates over metal-line cooling at temperatures of a few thousand Kelvin and below for
simulations with various different gas-phase metallicities and ultraviolet (UV) field strengths. However,
the limited spatial resolution of their model means that they have to include a clumping factor
into their H$_{2}$ formation rate in order to correct for unresolved small-scale density 
fluctuations. This correction factor is easy to justify within regions representing dense atomic
or molecular clouds, which are dominated by turbulence and therefore have significant 
density substructure, but it is less obvious whether including a correction of this form is
appropriate for regions corresponding to lower density, warm patches of the ISM, which
are dominated by thermal motions and hence have much less turbulent substructure.

Another important recent study is that of \citet{as11}. They performed a series of simulations
of a small protogalaxy ($M_{\rm tot} \sim 10^{9} \: {\rm M_{\odot}}$) and explored the effects
of varying the initial metallicity of the gas and the strength of the ambient radiation field. They
found that at metallicities below $10^{-2} \: {\rm Z_{\odot}}$, their model protogalaxies were cooled primarily by H$_{2}$
and that cooling in these low metallicity systems was very sensitive to the strength of the radiation
field, being largely suppressed for field strengths $G_{0} \geq 0.01$ in units of the \citet{habing68}
field.

One drawback of all of these studies is that their computational cost limits the number of simulations
that can be performed and makes it difficult to explore a wide parameter space of densities,
metallicities etc.\ Their results suggest that there are situations in which H$_{2}$ cooling is
important in the low density ISM, but these simulations do not, by themselves, allow us to 
identify the full range of physical conditions for which this is the case. For this reason, it is
useful to study the interplay between chemistry and thermodynamics in the low density ISM
using simpler methods that do not have a large computational cost. 

In this paper, we carry 
out such a study. We use a simple one-zone model of the ISM, in which the chemical and 
thermal evolution of gas that is initially hot and ionized is followed in detail, but where the
gas density is held constant. With this setup, we can follow the evolution over long periods
of time for minimal computational cost, allowing us to explore the behaviour of the gas for a 
wide range of different  densities, metallicities and UV field strengths.

\section{Numerical approach}
\label{sec:num}
Our aim is to establish the physical conditions in which gas that is initially in a warm, 
low-density state can cool significantly within a dynamical time, which we take to be a necessary
prerequisite for the formation of molecular clouds and, ultimately, stars. In order to do this, we make
use of a very simple one-zone model of the thermal and chemical evolution of the ISM. In this model,
we keep the gas density fixed, and simply track the evolution of the gas and dust temperatures, as
well as the chemical composition of the gas. We note that although gas that is able to cool will likely
also increase its density -- either by gravitational collapse or due to the effects of isobaric compression
from surrounding warmer material -- gas that is unable to cool is unlikely to change its density significantly
unless perturbed by some external force. Our model is therefore well-suited to identifying which gas 
can cool and which cannot, but will provide an incomplete picture of the behaviour of gas that does 
manage to cool.
 
We model the chemical and thermal evolution of the gas in the ISM using a simplified chemical
network coupled with a detailed atomic and molecular cooling function. Our chemical model
is based on three main sources. We take our treatment of the collisional gas-phase chemistry of
hydrogen, helium and deuterium from \citet{clark11}, and our treatment of the collisional gas-phase
chemistry of carbon, oxygen and silicon from \citet{gj07}. Rates for the key photochemical reactions
(e.g.\ H$^{-}$ photodetachment, H$_{2}$ photodissociation) are taken from \citet{glo10}.  Our model tracks 
the abundances of atomic and ionized carbon, oxygen and silicon, but does not include the formation of 
molecules containing these elements (e.g.\ CO, water). Our justification for ignoring these heavy molecules
is that at the densities and temperatures considered in this study, the cooling they provide per molecule is
no more than a factor of a few larger than the cooling per atom provided by atomic fine structure cooling
\citep{gj07}. Therefore, cooling from these heavy molecules will be important only when a significant fraction
of the available carbon and oxygen has been incorporated into molecules. We know that in the local ISM,
this occurs only in dense, well-shielded gas, and this gas cools rapidly, in much less than a free-fall time,
even when cooling from these molecules is neglected \citep{gc12a}. At lower metallicity, the dust extinction 
in the ISM will be lower and we therefore expect these heavy molecules to be even less important for 
determining whether the gas can cool. We therefore do not expect that this simplification 
will significantly affect our conclusions. We account for the formation of H$_{2}$ on dust 
grains using the standard \citet{hm79} prescription, but  do not include any other grain-surface 
chemistry in our model. Heating in our model comes primarily from the photoelectric effect  and cosmic 
ray heating,  with additional minor contributions from effects such as H$_{2}$ formation heating or ultraviolet
pumping of excited vibrational states of H$_{2}$. Cooling is provided by the electronic excitation
of atomic H, He and He$^{+}$, rotational and vibrational emission from H$_{2}$ and HD, fine
structure and metastable line emission from C, C$^{+}$, O, Si and Si$^{+}$, and thermal emission from 
dust. Metastable line emission represents a new addition to our cooling function and is treated using
the data given in Table~9 of \citet{hm89}.

The collisional rate coefficients that we adopt for the various chemical reactions in our model are for 
the most part the same as in \citet{gj07} (for the metals) and \citet{clark11} (for the H, He and D 
chemistry), but in a few cases we have updated the values,  in light of new experimental
or theoretical data. For the associative detachment reaction 
\begin{equation}
{\rm H^{-}} + {\rm H} \rightarrow {\rm H_{2}} + {\rm e^{-}}
\end{equation}
we now use the new rate coefficient measured by \citet{kreck10} in place of the older value referenced 
in \citet{clark11}. For the mutual neutralization reaction
\begin{equation}
{\rm H^{-}} + {\rm H^{+}} \rightarrow {\rm H + H},
\end{equation}
we use the rate coefficient given in \citet{cdg99}, which agrees well with recent theoretical \citep{sten09}
and experimental \citep{urb12} determinations.  We have also updated several of the rate
coefficients used in our cooling function. To model the contribution to the H$_{2}$ cooling rate made
by H$_{2}$-proton collisions, we now make use of the excitation rates recently calculated by  
\citet{hon11,hon12} for the transitions for which these are available, supplementing them with
data from \citet{ger90} and \citet{kr02} for those transitions for which newer data is not available.
The contribution of H$_{2}$-electron collisions is now modelled using the excitation rate coefficient
data given in \citet{yoon08}, in place of the much older data used in our previous treatment
\citep[see][]{ga08}. Finally, in our treatment of fine structure cooling now makes use of the new
excitation rates for C-H and O-H collisions computed by \citet{akd07} in place of the older values 
used in \citet{gj07}.

We consider gas that is initially warm ($T = 10000 \: {\rm K}$) and fully ionized, although in 
Section~\ref{res:ion} below we discuss the effects of starting with a reduced level of ionization. We explore the influence of 
three main free parameters: the gas density, the metallicity, and the strength of the interstellar radiation 
field (ISRF). We can describe the gas density in terms of the number density of hydrogen nuclei $n$, which
is related to the mass density by $\rho = 1.4 m_{\rm p} n$, where $m_{\rm p}$ is the proton mass, and
we have assumed a 10:1 ratio of hydrogen to helium, as is appropriate for the local ISM. To parameterize
the metallicity, we assume that at solar metallicity, the elemental abundances of C, O and Si relative
to hydrogen are the same as those measured in the warm neutral medium \citep[see e.g.][]{sem00}. 
At other metallicities, we assume that
the elemental abundances simply scale linearly with the total metallicity Z. We therefore have
\begin{eqnarray}
x_{\rm C, tot} & = & 1.41 \times 10^{-4} \left(\frac{{\rm Z}}{{\rm Z_{\odot}}}\right), \\
x_{\rm O, tot} & = & 3.16 \times 10^{-4} \left(\frac{{\rm Z}}{{\rm Z_{\odot}}}\right), \\
x_{\rm Si, tot} & = & 1.51 \times 10^{-5} \left(\frac{{\rm Z}}{{\rm Z_{\odot}}}\right),
\end{eqnarray}
where $x_{\rm C, tot}$, $x_{\rm O, tot}$, and $x_{\rm Si, tot}$ are the total fractional abundances of
these three elements. We also assume in most of our models that the dust-to-gas ratio, ${\cal D}$, 
scales linearly with metallicity, although in Section~\ref{res:dust} below we report the results of runs 
in which we adopted a steeper dependence of ${\cal D}$ on Z.

To model the ISRF, we adopt the spectral shape described in
\citet{dr78}  in the ultraviolet, and that from \citet{mmp83} at longer wavelengths.
We assume that the shape of the spectrum does not change as we change the normalization, 
$G_{0}$, where setting $G_{0} = 1$ gives us the original Draine and Mathis et al.\ 
normalizations. 
In some of our models, we include an approximate treatment of the effects of H$_{2}$ self-shielding.
This is modelled using a self-shielding function based on the work of \citet{db96}, but modified
according to the prescription in \citet{wghb11} in order to more accurately represent the effects of 
self-shielding in warm gas. As input to the self-shielding function, we need to provide an H$_{2}$
column density. We compute this as $N_{\rm H_{2}} = n_{\rm H_{2}} L_{\rm ss}$, where $n_{\rm
H_{2}}$ is the current H$_{2}$ number density and $L_{\rm ss}$ is a characteristic self-shielding
scale length. In our default set-up, we set $L_{\rm ss} = 0$, so that there is no self-shielding.
We explore the effects of adopting a non-zero value of $L_{\rm ss}$ in Section~\ref{res:h2ss}.

We adopt a cosmic ray ionization rate for atomic hydrogen given by $\zeta_{\rm H}
= 10^{-16} G_{0} \: {\rm s^{-1}}$. We include a dependence on $G_{0}$ to reflect the
fact that the cosmic rays ionization rate and the UV photodissociation rate would both
be expected to increase as the star formation rate increases. The ionization rates for 
our other chemical species (H$_{2}$, He, C, etc.) relative to $\zeta_{\rm H}$ are 
scaled as described in \citet{gj07}.

\section{Results}
\label{sec:res}
\subsection{ISM cooling without H$_{2}$}
\begin{figure*}
\includegraphics[width=0.33\textwidth]{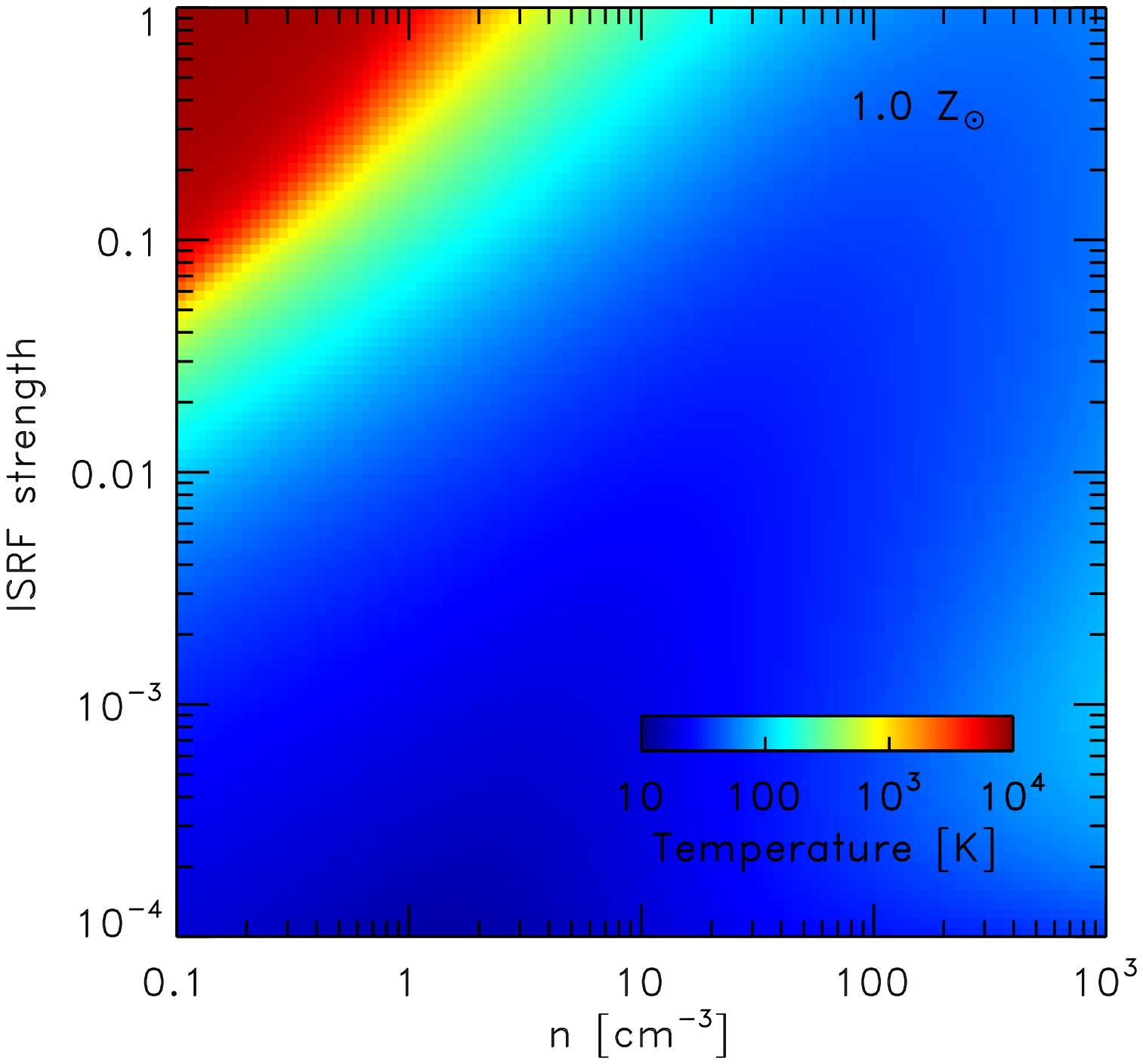}
\includegraphics[width=0.33\textwidth]{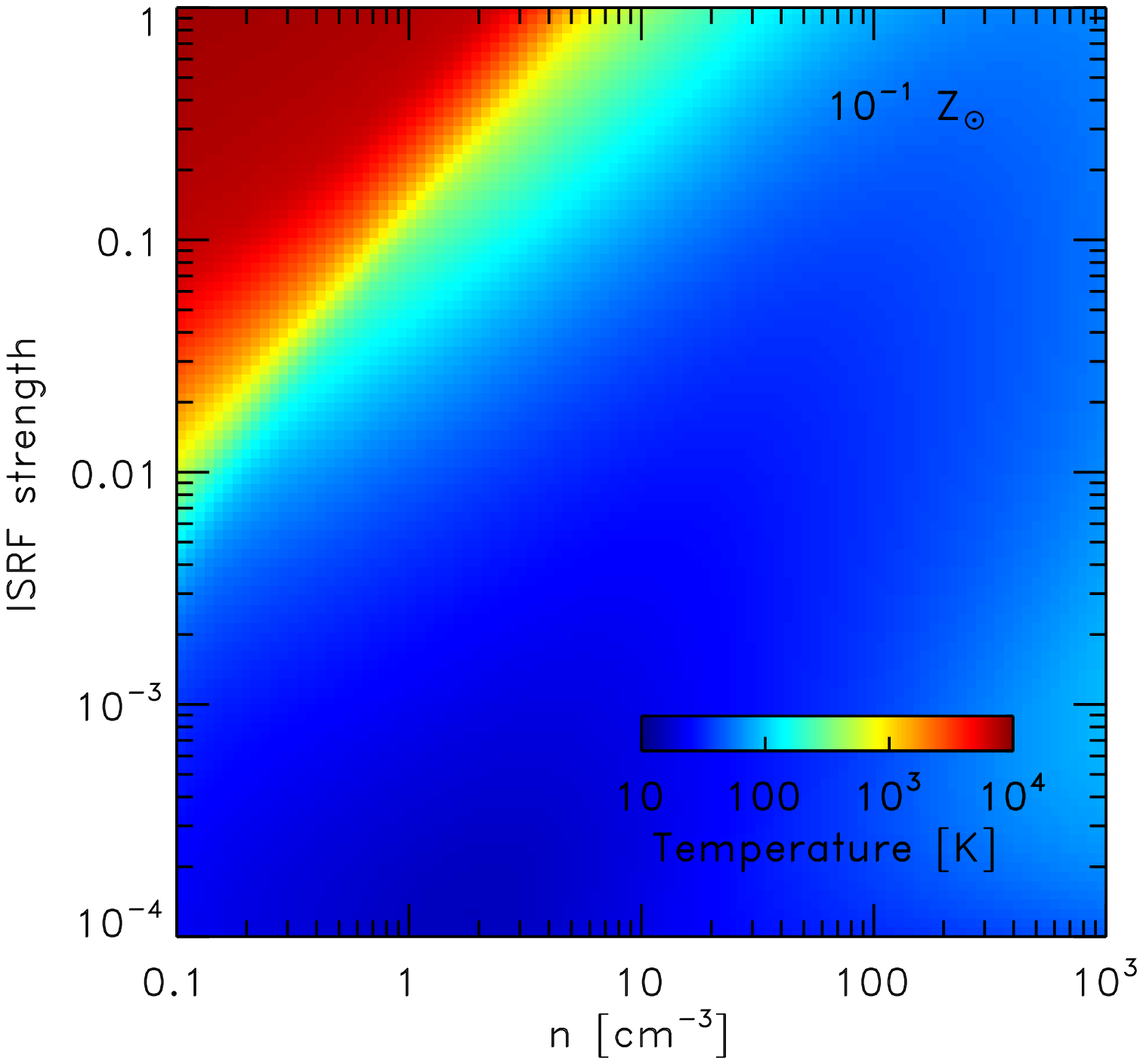}
\includegraphics[width=0.33\textwidth]{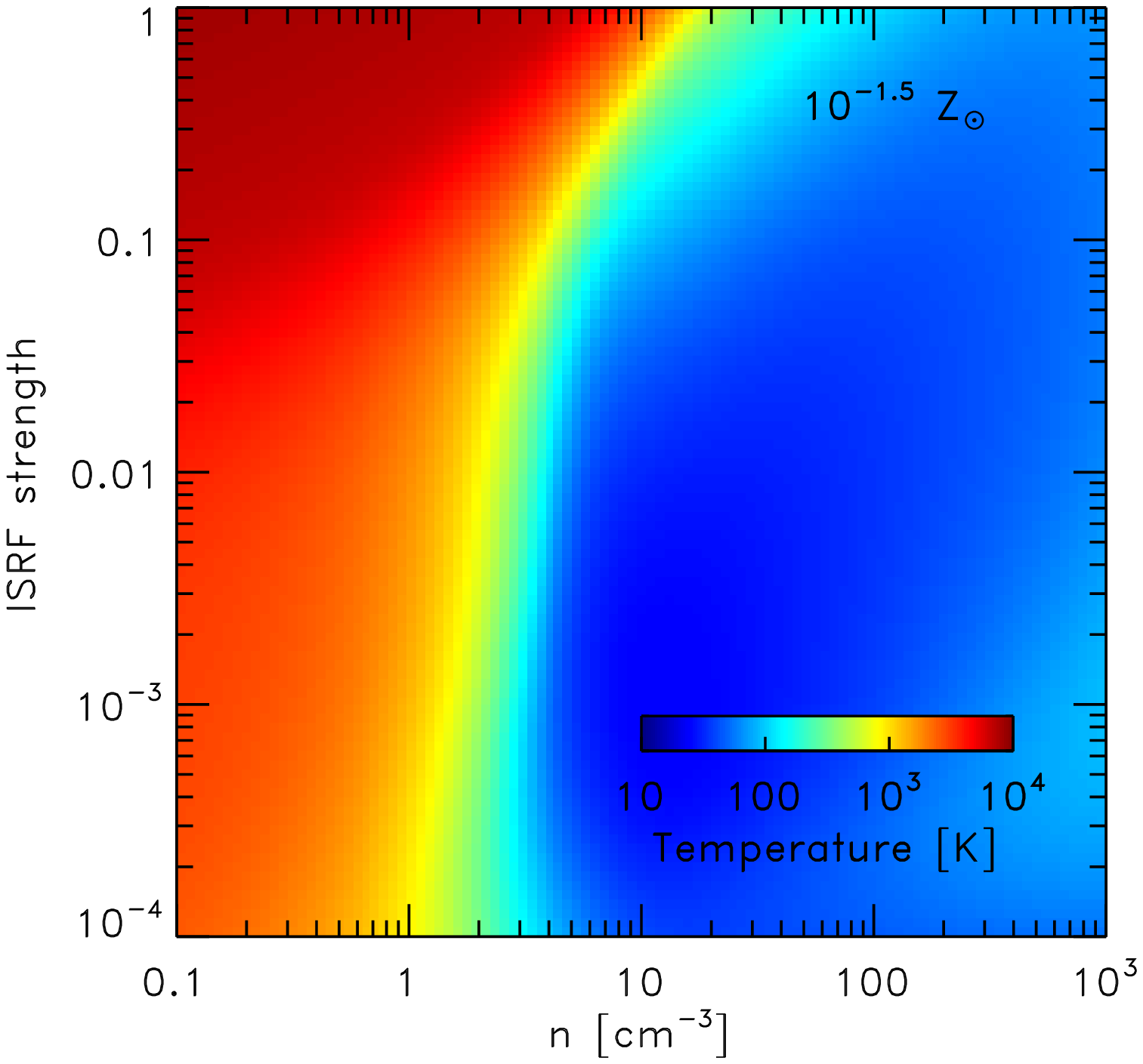}
\includegraphics[width=0.33\textwidth]{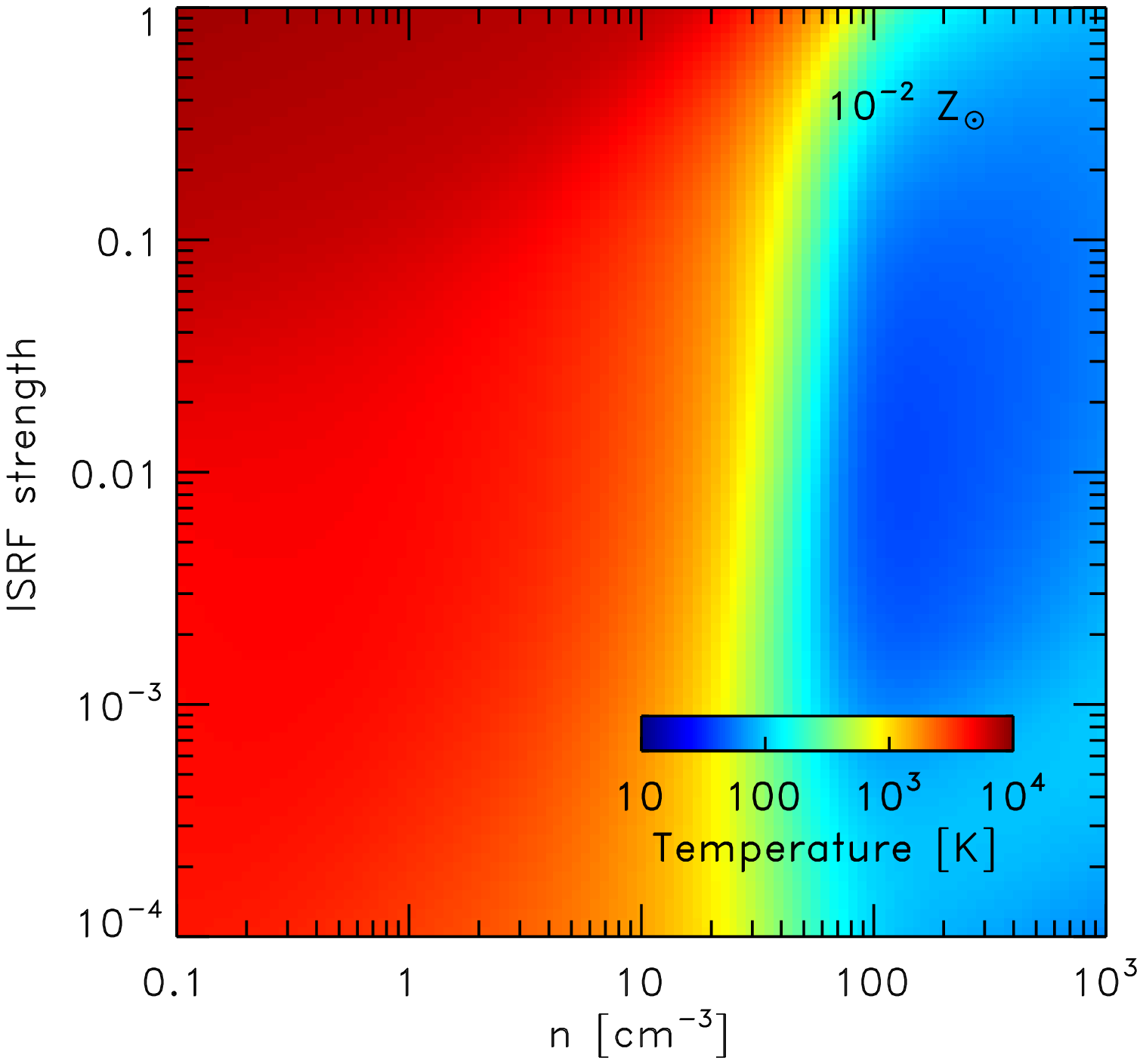}
\includegraphics[width=0.33\textwidth]{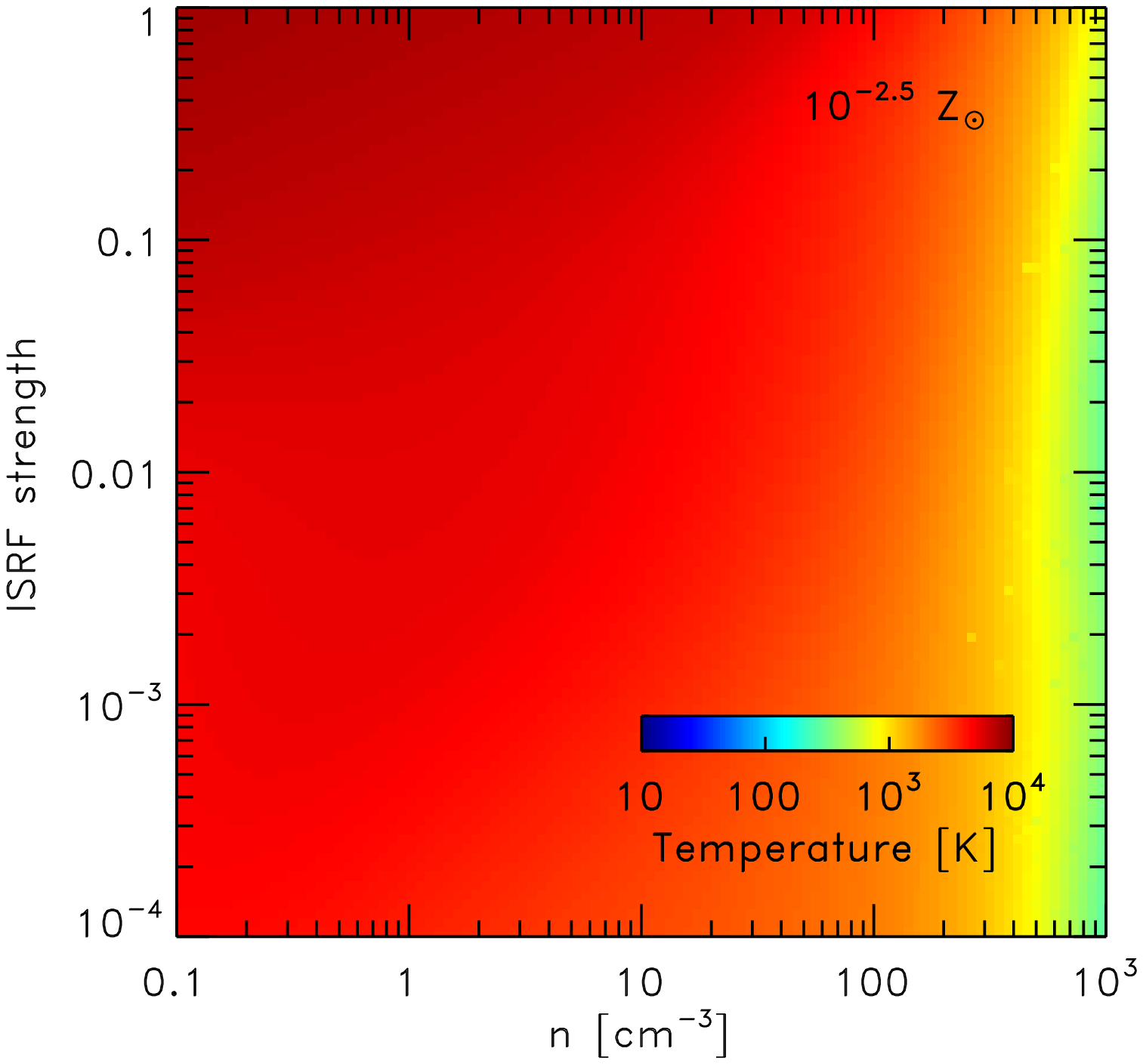}
\includegraphics[width=0.33\textwidth]{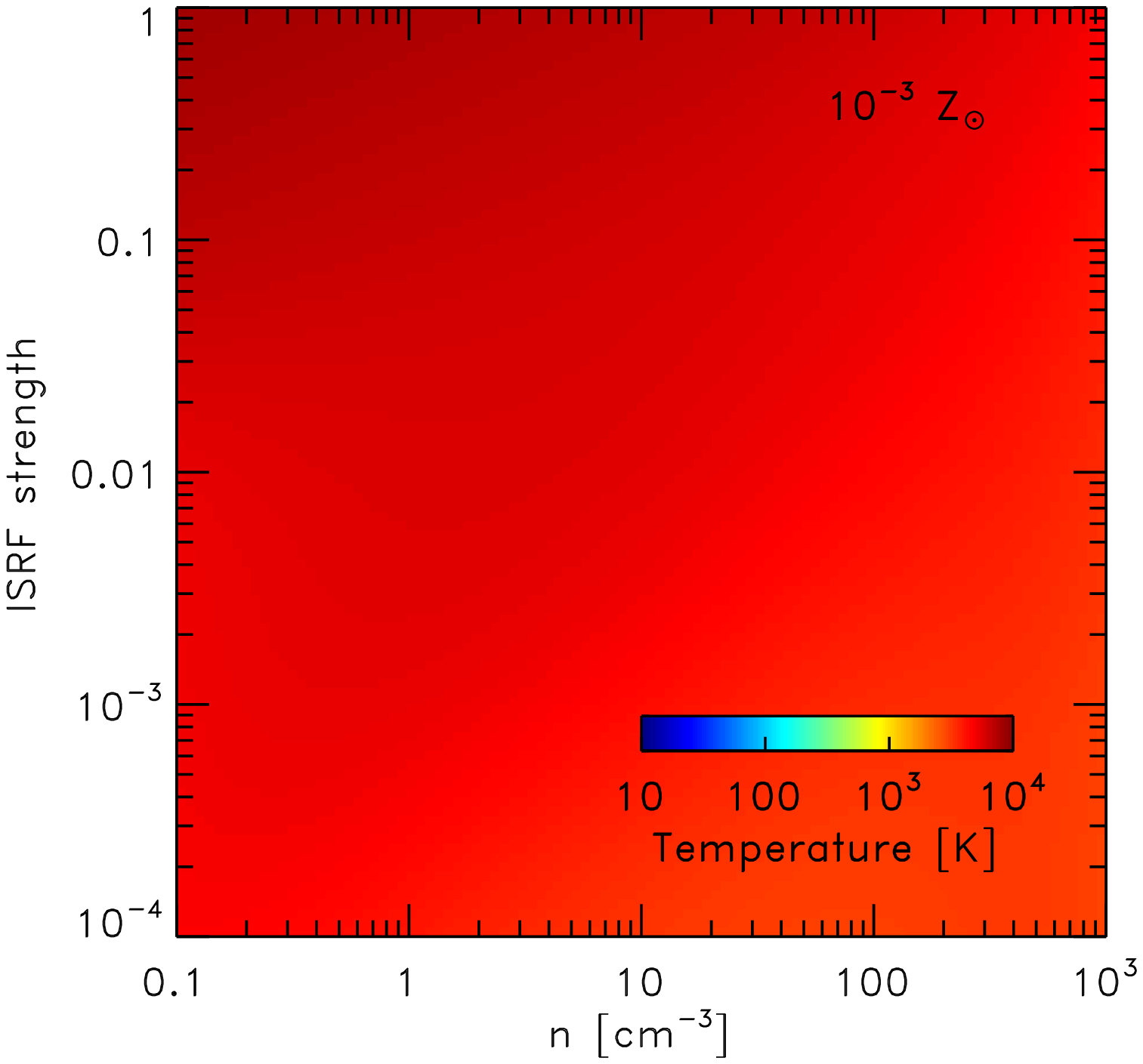}
\includegraphics[width=0.33\textwidth]{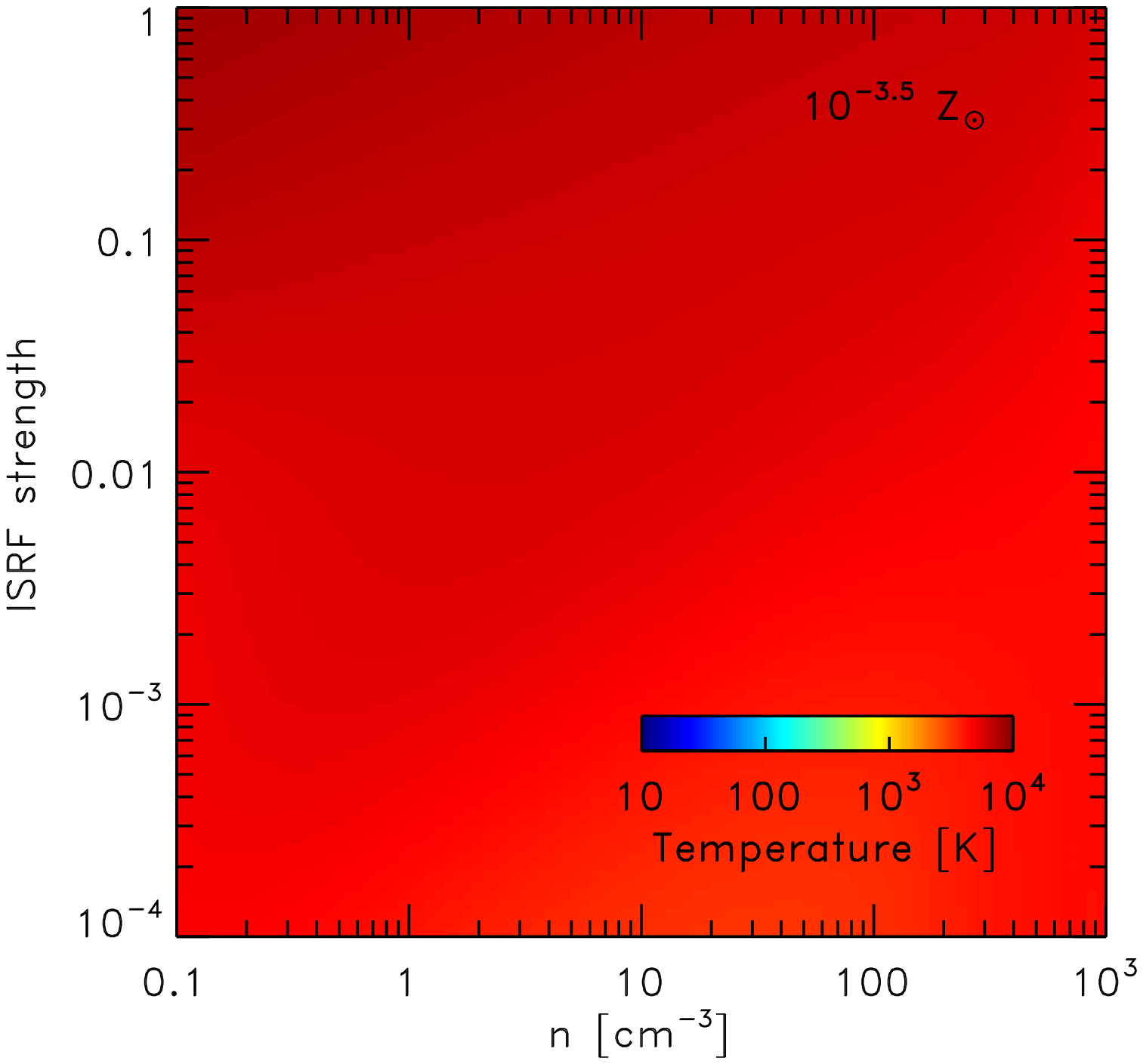}
\includegraphics[width=0.33\textwidth]{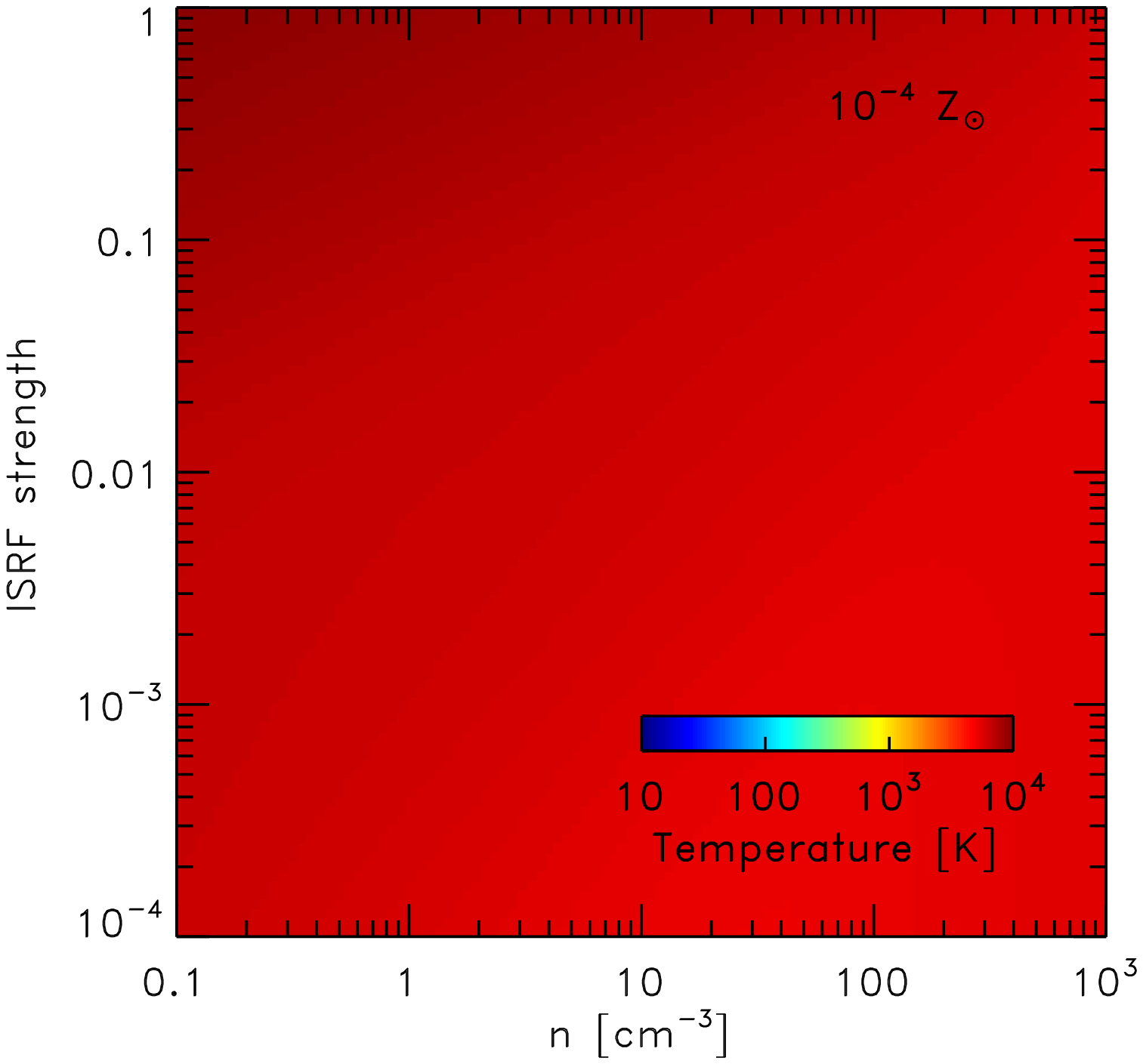}
\caption{Gas temperature at $t = t_{\rm ff}$, computed as a function of the
number density of hydrogen nuclei, $n$, and the strength of the 
interstellar radiation field in units of the standard value, $G_{0}$,
for a set of runs covering a range of metallicities between $Z = Z_{\odot}$
and $Z = 10^{-4} \: {\rm Z_{\odot}}$. In these runs, the effects of H$_{2}$ 
and HD cooling were not included. \label{no-H2}}
\end{figure*}

We begin our study by examining to what extent the ISM can cool in the absence of H$_{2}$ cooling.
To do this, we use the one-zone model described in the previous section to explore the thermal
evolution of the gas for a wide range of different  densities, $0.1 \leq n \leq 10^{3} \: {\rm cm^{-3}}$,
and ISRF strengths, $10^{-4} < G_{0} < 1.0$, at eight different metallicities: $\log \left({\rm Z} / {\rm Z_{\odot}}\right) = 
0.0, -1.0, -1.5, -2.0, -2.5, -3.0, -3.5$ and $-4.0$. For each value of Z, $G_{0}$ and $n$, we run the model 
for a single gravitational free-fall time, $t_{\rm ff} = (3\pi / 32 G \rho)^{1/2}$, and examine the temperature 
at the end of this time period. Our choice of the free-fall time here is motivated by the classic Rees-Ostriker
criterion for dynamical fragmentation \citep{ro77}. However, we have verified that our results are not significantly
different if we allow the gas to evolve for e.g.\ two free-fall times. Our results are plotted in Figure~\ref{no-H2}.

At solar metallicity, we see that there are two main temperature regimes. For values of $G_{0} / n$ greater than
around 0.3,  the gas  remains hot, with a temperature of around 8000--9000~K. For lower values of $G_{0} / n$, 
however, the gas cools significantly, reaching temperatures of order a few hundred K when $G_{0} / n \sim 0.3$, 
and temperatures as low as 20~K when $G_{0} / n \ll 0.3$. We can understand this behaviour as a consequence 
of the fact that over a wide range in temperatures, the cooling of the neutral ISM is dominated by fine structure line 
emission, while the heating is dominated by photoelectric heating \citep[see e.g.][]{w95,w03}. If we write the cooling 
rate as 
\begin{equation}
\Lambda_{\rm fs} = \Lambda_{0}(T) \frac{{\rm Z}}{{\rm Z_{\odot}}} n^{2},
\end{equation}
where $\Lambda_{0}$ depends only weakly on $T$ for $T > 100$~K, and the heating rate as 
 \begin{equation}
\Gamma_{\rm pe} = \Gamma_{0}(n_{\rm e}, T) \frac{{\cal D}}{{\cal D_{\odot}}} G_{0} n,
\end{equation}
where  $\Gamma_{0}$ is also a weak function of $T$, then it is easy to show that in solar metallicity gas,
\begin{equation}
\frac{\Gamma_{\rm pe}}{\Lambda_{\rm fs}} \propto \frac{G_{0}}{n}.
\end{equation}
Since the temperature dependence of both the photoelectric heating rate and the fine structure cooling rate is 
weak in the temperature range of interest, it is the size of the factor $G_{0} / n$ that primarily determines whether 
heating or cooling dominates, and hence whether the gas remains close to its starting temperature, or cools 
until it reaches a temperature at which further fine structure cooling becomes ineffective. 

If we now decrease the metallicity of the gas by an order of magnitude, we see that the main change which
occurs is a slight increase in the size of the region in which cooling is ineffective. The critical value of 
$G_{0} / n$ decreases from 0.3 to around 0.1, but the behaviour of the gas for $G_{0} / n \ll 0.1$ or
$G_{0} / n \gg 0.1$ does not significantly change. Again, this can be understood
as a consequence of the cooling rate being dominated by fine structure emission and the heating rate
by photoelectric emission from dust. Since we have assumed that the dust-to-gas ratio is proportional
to the metallicity, both the heating and the cooling rates scale linearly with the metallicity, and the
resulting equilibrium temperature is largely independent of metallicity. The fact that it is not completely
independent is largely due to the fact that the cosmic ray heating rate does not decrease when the 
metallicity decreases, meaning that cosmic ray heating becomes increasingly dominant as we move
to lower metallicities.

If we decrease the metallicity further, to ${\rm Z} = 10^{-1.5} \: {\rm Z_{\odot}}$, we find that there is a
qualitative change in the behaviour of the gas.  At densities $n > 3  \: {\rm cm^{-3}}$, the gas behaves
largely as before, but at lower densities, the gas remains warm regardless of the value of $G_{0}$. 
This change in behaviour is not due to a change in the equilibrium temperature of the gas, which
remains low when $G_{0}$ is low. Instead, it reflects the fact that at low densities in this simulation,
the gas does not have time to reach its equilibrium temperature. In this regime, the cooling time
 \begin{equation}
t_{\rm cool} = \frac{1}{\gamma - 1} \frac{n_{\rm tot} kT}{\Lambda},
\end{equation}
becomes larger than the free-fall time $t_{\rm ff}$ before the temperature of the gas can reach
equilibrium. For example, consider the case of gas with $T = 6000$~K at a number
density $n = 0.1 \: {\rm cm^{-3}}$. For these conditions, and for ${\rm Z} = 10^{-1.5} \: {\rm Z_{\odot}}$,
the total cooling rate due to fine structure emission  from C$^{+}$, Si$^{+}$ and O is approximately 
$\Lambda_{\rm tot} \simeq 9.4 \times 10^{-30} \: {\rm erg} \: {\rm s^{-1}} \: {\rm cm^{-3}}$. The cooling
time of this gas is therefore at least $1.3 \times 10^{16} \: {\rm s}$, regardless of the value of $G_{0}$.
This is significantly longer than the free-fall time, which for gas at this density is 
approximately $t_{\rm ff} \simeq 4 \times 10^{15} \: {\rm s}$, and hence the gas is unable to cool
below around 6000~K within a free-fall time.

Continuing to decrease the metallicity beyond this point makes it even harder for the gas to cool
within a free-fall time, as its cooling time scales inversely with metallicity while its free-fall time remains
fixed. By the time we reach a metallicity of ${\rm Z} = 10^{-3} \: {\rm Z_{\odot}}$, we see that none of
the gas in the region of G$_{0}$--$n$ parameter space examined here can cool very much within a
free-fall time.

It is also easy to see how these results would change if we examined the gas after
a longer period of time. Since the limiting factor in the gas with low density and low $G_{0}$ is
the time available for cooling, increasing the time period considered from $t_{\rm ff}$ to e.g.\
$2 t_{\rm ff}$ would shift the density at which the transition from hot gas to cold gas occurs,
in this case by a factor of 4. Our precise results therefore depend on our choice of $t_{\rm ff}$
as the moment to examine the gas temperature. However, to completely offset the effects of
a substantial drop in metallicity, one would have to consider a much longer time period; for
example, a decrease in ${\rm Z_{\odot}}$ by a factor of ten could be mitigated by looking at
the gas after ten free-fall times. In practice, it is unlikely that the gas would remain in an
undisturbed state for such a long time period, and even if it did so, gas which can cool and
collapse only on a timescale $t \gg t_{\rm cool}$ is unlikely to form stars efficiently.

In summary, we see that for the majority of the metallicities considered here, the main factor that 
limits the temperature to which the gas can cool within a free-fall time is not the influence of 
photoelectric and/or cosmic ray heating, it is simply the fact that the cooling time is long 
compared to the free-fall time.  

\subsection{ISM cooling with H$_{2}$}
\label{sec:H2}
\begin{figure*}
\includegraphics[width=0.33\textwidth]{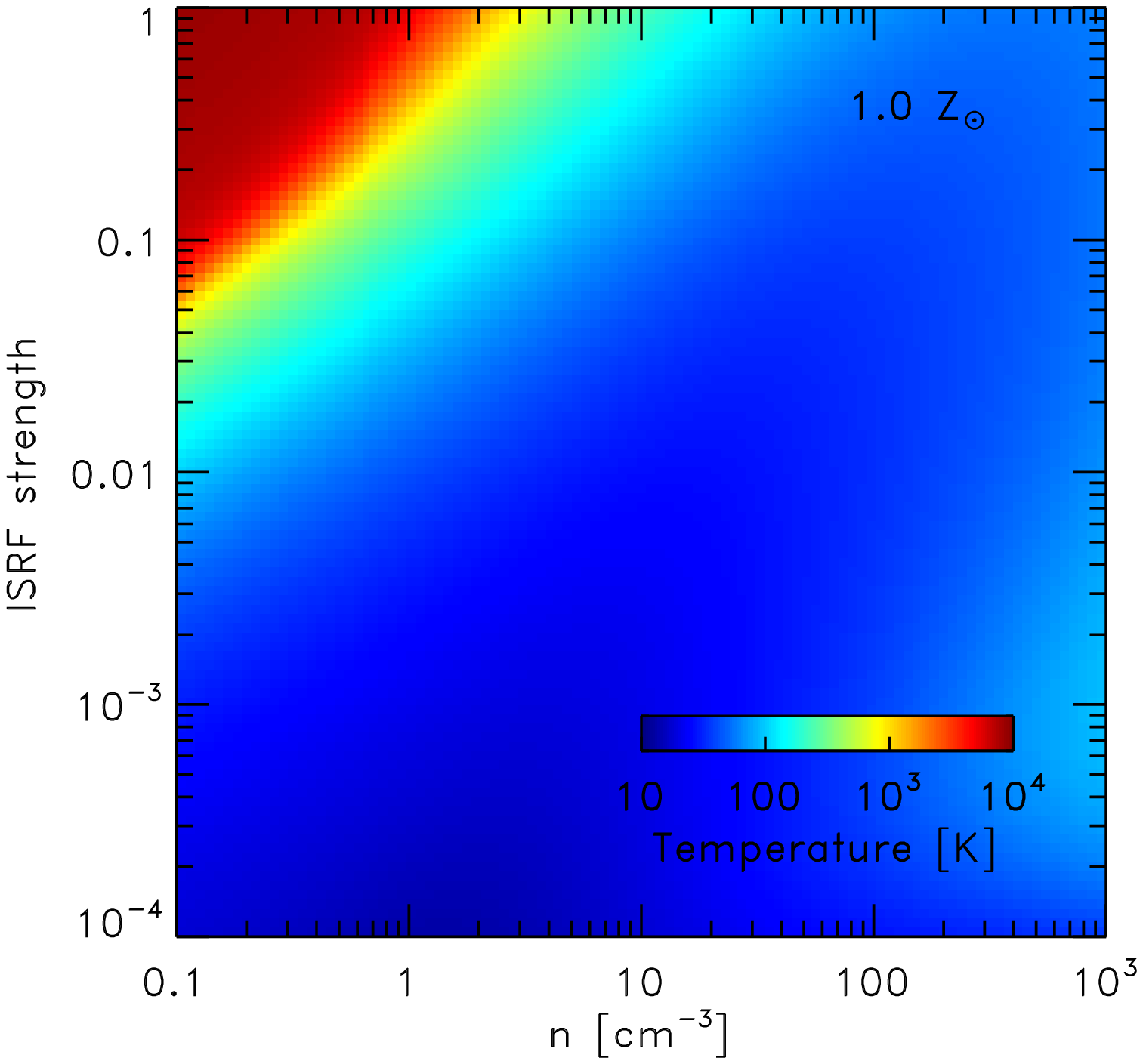}
\includegraphics[width=0.33\textwidth]{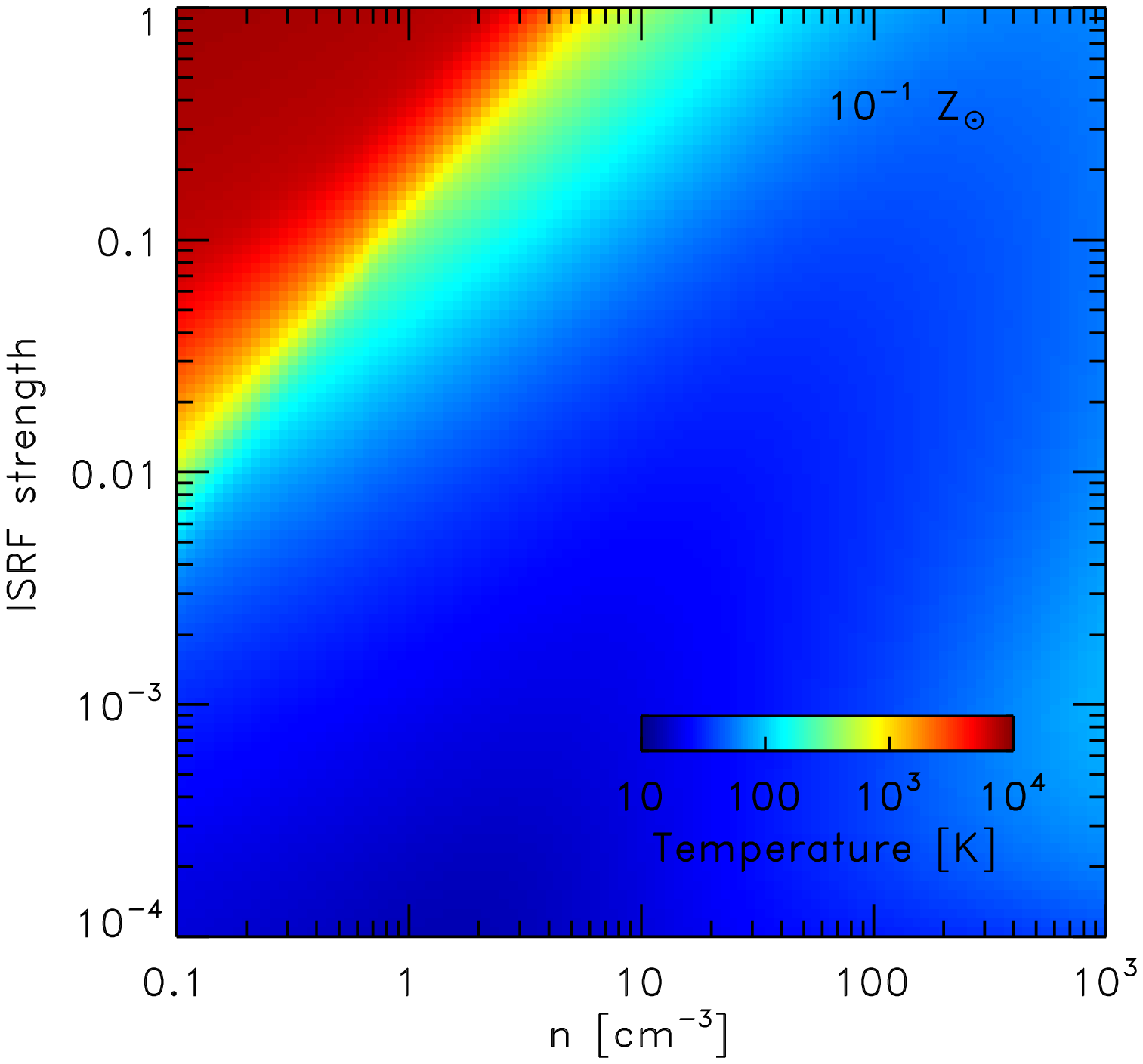}
\includegraphics[width=0.33\textwidth]{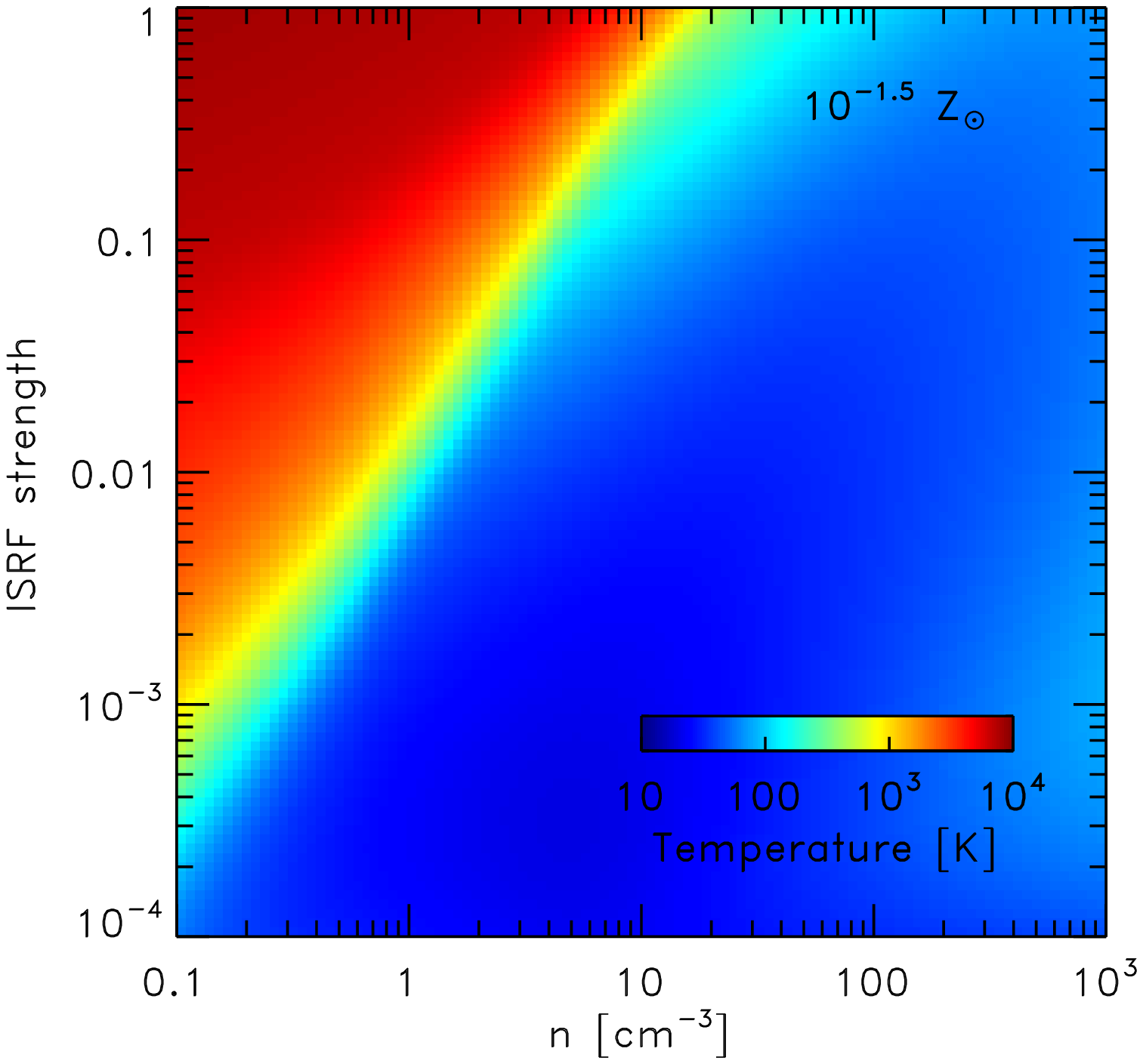}
\includegraphics[width=0.33\textwidth]{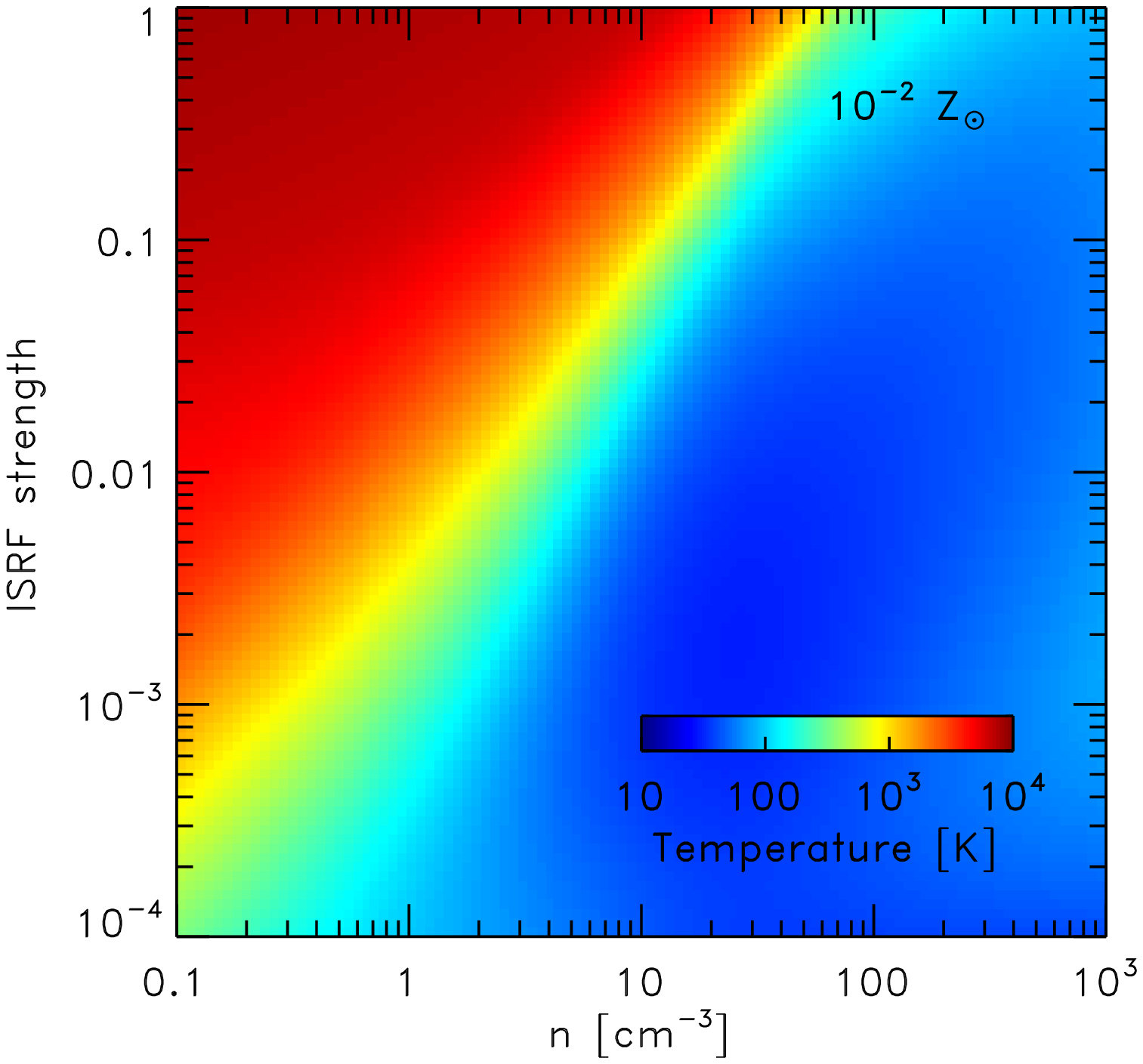}
\includegraphics[width=0.33\textwidth]{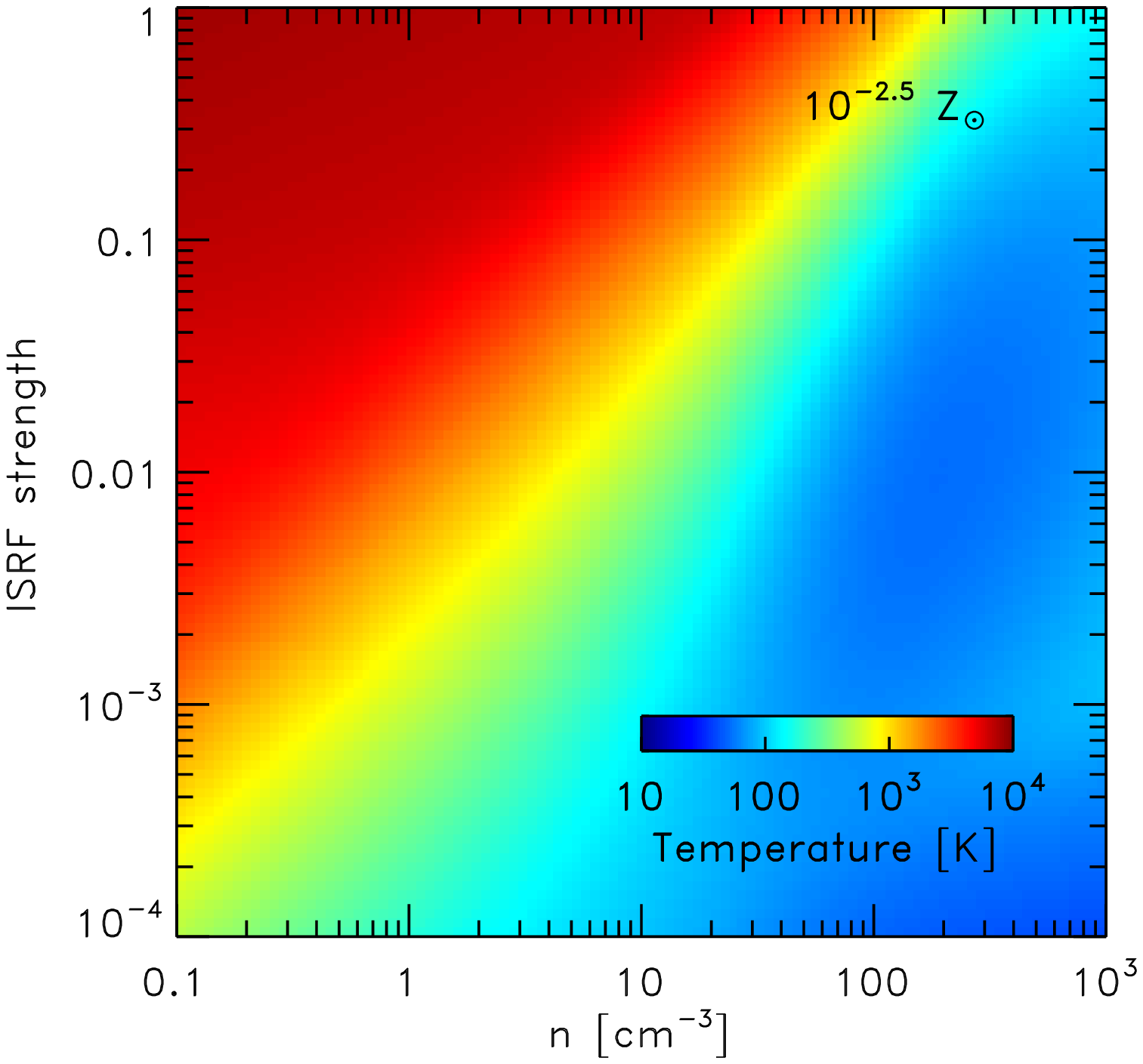}
\includegraphics[width=0.33\textwidth]{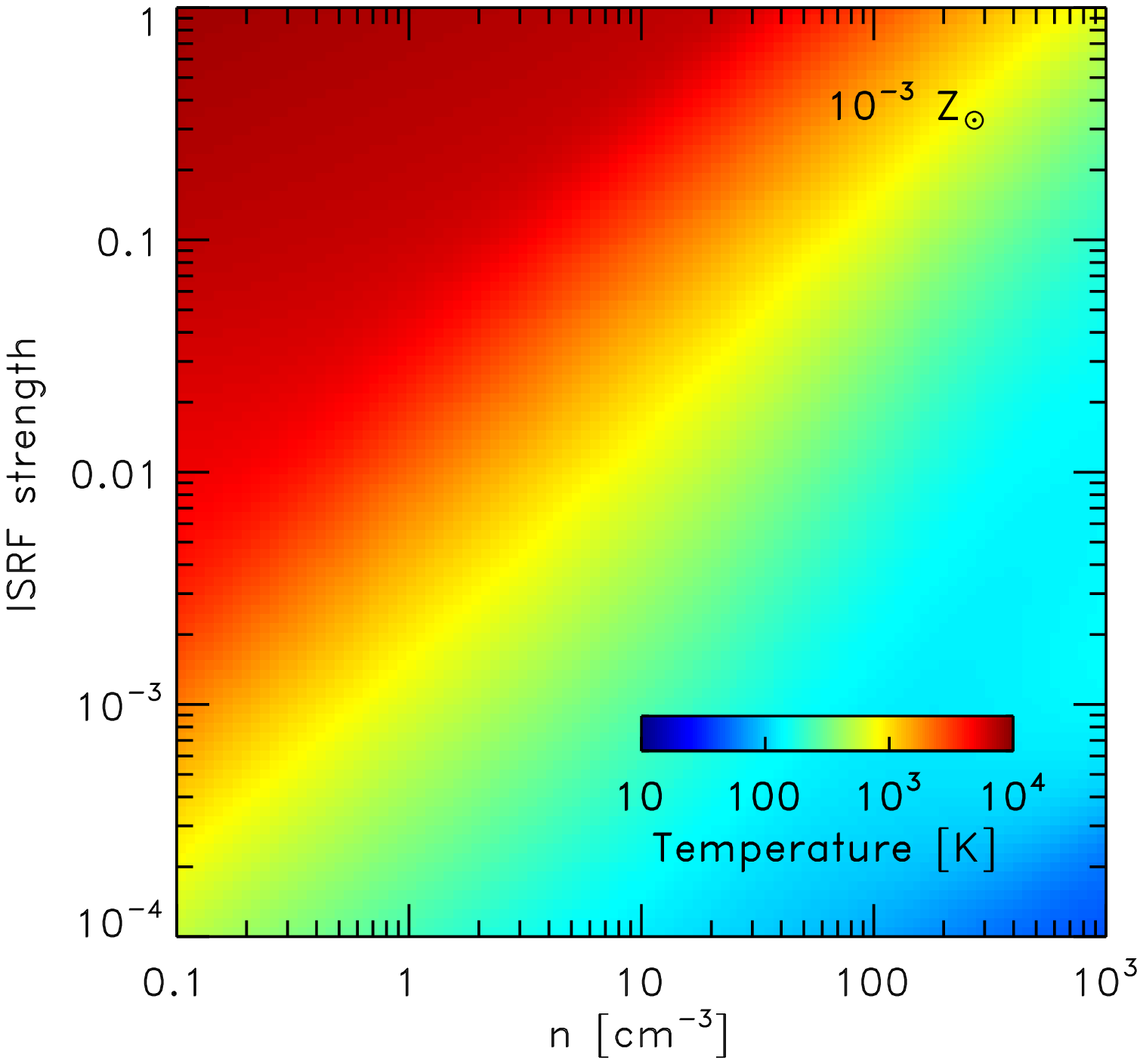}
\includegraphics[width=0.33\textwidth]{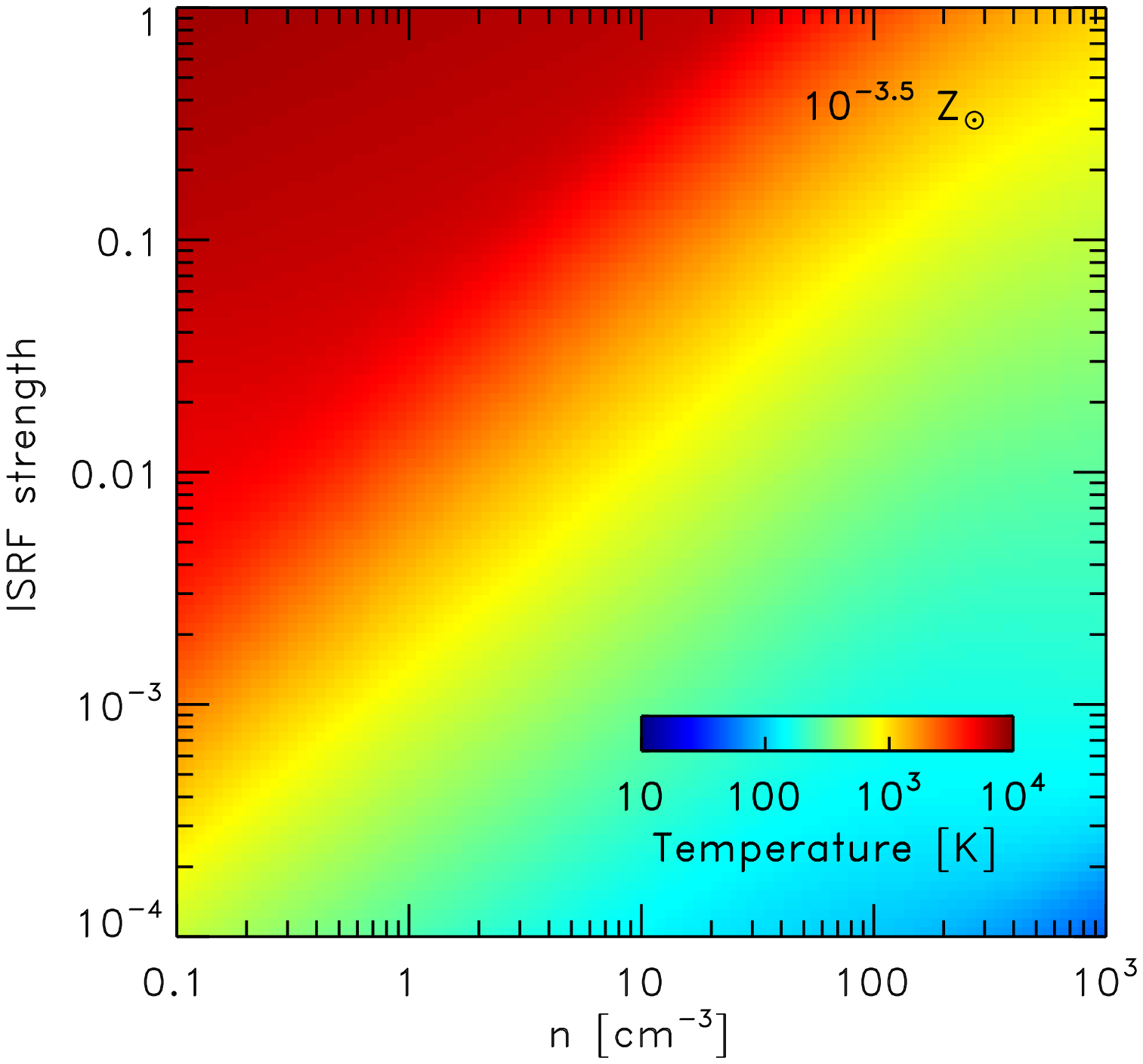}
\includegraphics[width=0.33\textwidth]{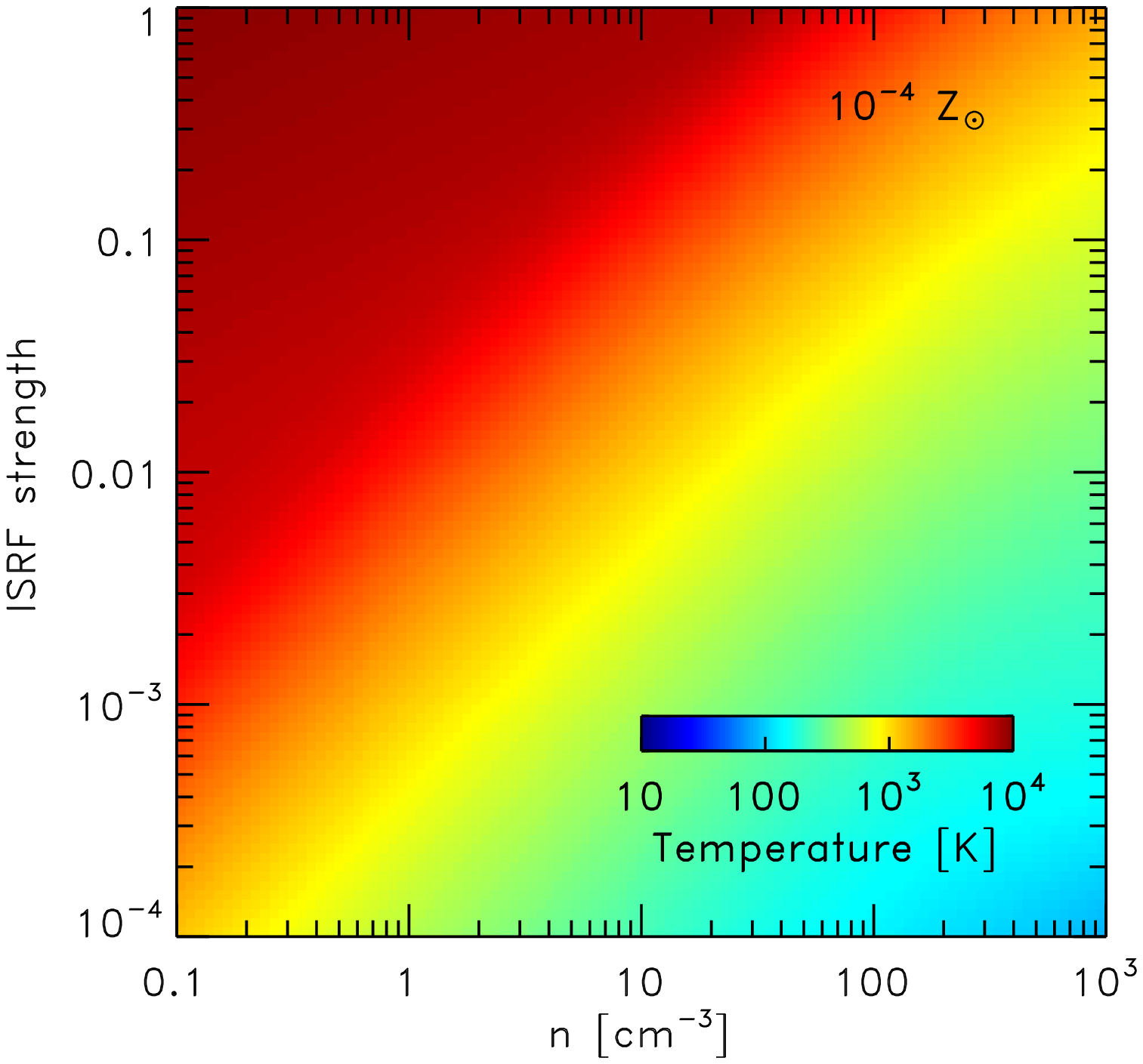}
\caption{As Figure~\ref{no-H2}, but for a set of runs that included
the effects of H$_{2}$ and HD cooling. \label{with-H2}}
\end{figure*}

We next look at how the ability of the ISM to cool changes once we include the effects of H$_{2}$ and HD cooling.
In Figure~\ref{with-H2}, we show the results of a similar set of simulations to those in the previous section. The
only difference in these simulations is that in this case, we account for molecular cooling. We see immediately that
at solar metallicity, there is essentially no difference in the outcome. Indeed, if we quantitatively compare the temperatures
reached at each point in $G_{0}$--density space in this run with those in the model without H$_{2}$ cooling, we find
that the maximum difference in the final temperature is less than 1\%. We have also looked in detail at the fraction of
the total cooling provided by H$_{2}$ throughout the lifetime of each of the runs, and find that this is never greater
than around 10\%. We can therefore immediately conclude that in metal-rich gas, and in the absence of self-shielding,
H$_{2}$ cooling does not play a significant role in regulating the temperature. This is consistent with the idea that in 
galaxies like our own Milky Way, the observed correlation between molecular gas and star formation is a consequence 
of the fact that both molecules and stars form preferentially in regions with high column density and high volume density, 
rather than an indication that molecular cooling is required for star formation  \citep[see e.g.][]{klm11,gc12a,gc12b,krum12}.

At lower metallicities, however, we start to see significant differences between the two sets of runs. 
We have already seen that in the runs without molecular cooling, as we lower the metallicity, the
the question of whether a given parcel of gas can cool within a free-fall time becomes determined
almost entirely by the gas density, with $G_{0}$ having little or no influence on whether the gas can
cool. This is not the case in the runs that include molecular cooling. In these runs, $G_{0}$ plays an
important role in determining the outcome of the simulations for all of the metallicities that we examine.
For metallicities in the range $10^{-1.5} > {\rm Z} > 10^{-2.5} \: {\rm Z_{\odot}}$, the boundary dividing those 
regions in  $G_{0}$--density space that cool from those that do not roughly follows a line along which  
$G_{0} / n^{3/2}$ is constant, while at lower metallicities, the dividing line is better described as a line 
of constant $G_{0} / n$. We see also that the temperature reached by the cooling gas increases significantly 
as we decrease the metallicity, from $T \sim 10$--20~K at solar metallicity to $T \sim 200$--300~K at 
$10^{-4} \: {\rm Z_{\odot}}$, although we caution that these numbers should be treated with care, as in
a realistic system, it is likely that the gas would not remain at constant density as it cools.
The difference between the runs with and without molecular cooling is particularly apparent at metallicities 
below $10^{-2} \: {\rm Z_{\odot}}$. At these metallicities, little cooling occurs in the runs without H$_{2}$,
whereas in the runs with H$_{2}$, cooling remains efficient over a large portion of the parameter space
that we examine.

In order to understand the behaviour that we find in these runs, and in particular why we recover the
scalings that we do, it is useful to look at how the fractional abundance of H$_{2}$ evolves with time in gas. 
At metallicities ${\rm Z} > 0.01 \: {\rm Z_{\odot}}$, the final H$_{2}$ fraction is set by the balance between H$_{2}$ 
formation on dust and H$_{2}$ photodissociation \citep{glo03}. We can write the H$_{2}$ formation rate as
\begin{equation}
R_{\rm form} = k_{\rm form} \, n \, n_{\rm H},
\end{equation}
where $k_{\rm form}$ is the reaction rate coefficient for H$_{2}$ formation on grain surfaces \citep[see e.g.][]{hm79,cs04},
$n$ is the number density of hydrogen nuclei and $n_{\rm H}$ is the number density of atomic hydrogen. 
For the H$_{2}$ photodissociation rate, we have 
\begin{equation}
R_{\rm dis} = k_{\rm dis} n_{\rm H_{2}},
\end{equation}
where $k_{\rm dis}$ is the photodissociation rate per H$_{2}$ molecule, which for our adopted radiation field is
given by $k_{\rm dis} = 5.6 \times 10^{-11} G_{0} \: {\rm s^{-1}}$. Combining these equations, it is easy to show that
in  chemical equilibrium, the H$_{2}$ number density is simply
\begin{equation}
n_{\rm H_{2}, eq} = \frac{k_{\rm form}}{k_{\rm dis}} \, n \, n_{\rm H} \propto G_{0}^{-1} n^{2}. \label{nh2_eq_hz}
\end{equation}
This equilibrium state is reached on a timescale $t_{\rm dis} = k_{\rm dis}^{-1} \simeq 1.8 \times 10^{10} G_{0}^{-1} 
\: {\rm s}$, and hence $t_{\rm dis} \ll t_{\rm ff}$ for the majority of the densities and UV field strengths that we examine. 
At densities far below the H$_{2}$ critical density $n_{\rm crit} \sim 10^{4} \: {\rm cm^{-3}}$,
the cooling time of the gas due to H$_{2}$ line cooling can be written as
\begin{equation}
t_{\rm cool} \propto \frac{nT}{\Lambda_{\rm H_{2}} n n_{\rm H_{2}}},
\end{equation}
where $\Lambda_{\rm H_{2}}$ is the H$_{2}$ cooling rate expressed in units of ${\rm cm^{3}} \: {\rm s^{-1}}$,
which is in general a strong function of temperature. At a fixed gas temperature, $t_{\rm cool}$ therefore scales with 
the H$_{2}$ number density as
\begin{equation}
t_{\rm cool} \propto n_{\rm H_{2}}^{-1}.
\end{equation}
If we now set $t_{\rm cool} = t_{\rm ff}$, then since $t_{\rm ff} \propto n^{-1/2}$, it is easy to demonstrate that our
required H$_{2}$ number density varies as
\begin{equation}
n_{\rm H_{2}, req} \propto n^{1/2}
\end{equation}
as we change $n$. Comparing this with Equation~\ref{nh2_eq_hz}, we see that if for some combination of $n$ and
$G_{0}$ we have $n_{\rm H_{2}, eq} = n_{\rm H_{2}, req}$ (i.e.\ our equilibrium H$_{2}$ fraction is just enough to
cool the gas within a free-fall time), then the same condition will hold for other values of $G_{0}$ and $n$ only if
\begin{equation}
\frac{G_{0}}{n^{3/2}} \simeq {\rm constant},
\end{equation}
in agreement with the behaviour that we see in Figure~\ref{with-H2}.

\begin{figure}
\includegraphics[width=0.5\textwidth]{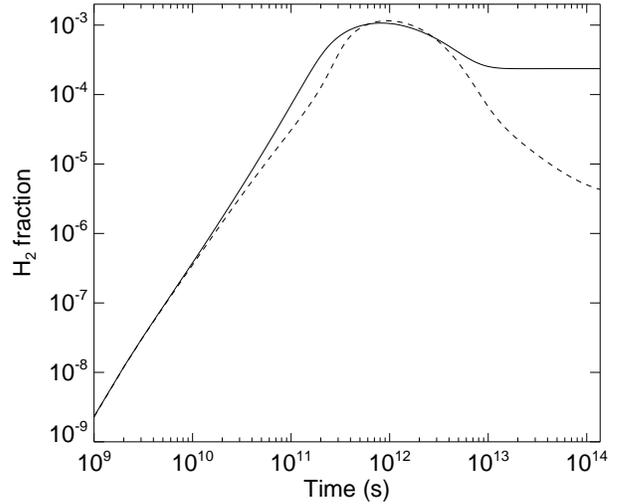}
\caption{Fractional abundance of H$_{2}$, plotted as a function of time, in gas with density $n = 100 \: {\rm cm^{-3}}$,
illuminated by an ISRF with strength $G_{0} = 0.01$, and with metallicity  ${\rm Z} = 0.1 \: {\rm Z_{\odot}}$ 
(solid line) or ${\rm Z} = 10^{-3} \: {\rm Z_{\odot}}$ (dashed line). \label{time-evol}}
\end{figure}

Below ${\rm Z} \sim 0.01 \: {\rm Z_{\odot}}$, the behaviour of the gas changes, owing to the increasing influence of the
gas-phase processes 
\begin{eqnarray}
{\rm H} + {\rm e^{-}} & \rightarrow & {\rm H^{-}} + \gamma, \label{ra} \\
{\rm H} + {\rm H^{-}} & \rightarrow & {\rm H_{2}} + {\rm e^{-}}. \label{ad}
\end{eqnarray}
The behaviour of the gas in this regime has been analyzed in detail by \citet{oh02}, who show that the ability of the gas
to cool is determined by the size of the ratio $G_{0} / n$. When $G_{0} / n$ is very small 
($G_{0} / n \ll 10^{-4}$)\footnote{In studies of the effects of UV radiation in low metallicity systems in the early Universe,
it is more common to quantify the strength of the UV radiation background in terms of $J_{21}$, the average UV flux in
the Lyman-Werner bands in units of $10^{-21} \: {\rm erg \: s^{-1} \: cm^{-2} \: Hz^{-1} \: sr^{-1}}$. To convert approximately
between these two different normalization schemes, we can use the fact that $G_{0} = 1$ corresponds to $J_{21} \simeq 40$.},
the most important H$_2$ destruction process is the charge transfer reaction
\begin{equation}
{\rm H_{2}} + {\rm H^{+}} \rightarrow {\rm H_{2}^{+}} + {\rm H}.  \label{h2_dest_ct}
\end{equation}
Some of the H$_{2}^{+}$ ions produced by this reaction reform H$_2$ via the inverse reaction
\begin{equation}
{\rm H_{2}^{+}} + {\rm H} \rightarrow {\rm H_{2}} + {\rm H^{+}},
\end{equation}
but many are instead destroyed by dissociative recombination,
\begin{equation}
{\rm H_{2}^{+}} + {\rm e^{-}} \rightarrow {\rm H} + {\rm H},
\end{equation}
and the end result is a net loss of H$_2$ from the gas. At high temperatures, this process strongly limits the equilibrium
H$_{2}$ abundance. However, as the gas cools, the rate of reaction~\ref{h2_dest_ct} falls off exponentially, allowing
more H$_{2}$ to form (and more cooling to occur), until the H$_2$ fractional abundance freezes out at a value of
approximately $10^{-3}$ \citep{teg97,oh02}, owing to the ongoing loss of electrons from the gas.

When $G_{0} / n \geq 10^{-4}$, on the other hand, the behaviour of the H$_{2}$ fraction is quite different. Initially, it increases  
steadily with time until it reaches an equilibrium set by the balance between H$_2$ formation via H$^{-}$ and H$_{2}$
destruction by charge transfer and by photodissociation. Following this, however, it begins to decrease again as the
electron fraction drops and the gas recombines. The falling electron fraction leads to a drop in the H$_{2}$ formation rate, 
and although the destruction rate due to charge transfer also falls, the photodissociation rate does not, and so the
equilibrium H$_{2}$ fraction decreases. 

This behaviour is illustrated in Figure~\ref{time-evol}, where we plot the time 
evolution of the H$_{2}$ fraction for two different metallicities, ${\rm Z} = 0.1 \: {\rm Z_{\odot}}$  and  
${\rm Z} = 10^{-3} \: {\rm Z_{\odot}}$, for gas with $n = 100 \: {\rm cm^{-3}}$ and $G_{0} = 0.01$. In the higher metallicity
run, the H$_{2}$ fraction increases steadily with time until it reaches a peak value of around $10^{-3}$ at $t \sim 10^{12}
\: {\rm s}$. At this point, the fractional ionization of the gas is around $x \sim 7.5 \times 10^{-3}$, H$_{2}$ formation is dominated 
by the gas-phase reaction chain, and the equilibrium H$_{2}$ fraction is around $10^{-3}$. At later times, the equilibrium
H$_{2}$ fraction decreases as the gas recombines. However, once the fractional ionization has decreased by a further
order of magnitude, H$_{2}$ formation on grain surfaces takes over as the most important process, and the H$_{2}$
fraction tends to a roughly constant equilibrium value of $x_{\rm H_{2}} \sim 2 \times 10^{-4}$. In the lower metallicity
run, the initial behaviour is very similar. However, in this case, the much lower dust abundance means that H$_{2}$
formation on grain surfaces is far less effective and gas-phase formation of H$_{2}$ remains the dominant H$_2$
formation process throughout the period shown in the figure. As a result, the H$_2$ fraction is far more sensitive to
the falling fractional ionization of the gas, and by the end of the period plotted has decreased by almost two orders of
magnitude in comparison to its peak value.

The significant fall-off with time of the H$_{2}$ fraction that occurs in the low metallicity run, together with the fact that the 
decreasing  ionization also leads to a decreasing cooling rate per H$_2$ molecule, implies that most of the cooling that
occurs in this case takes place at early times, prior to or at the point at which the H$_{2}$ abundance reaches its peak
value. Consequently, the important timescale in this case is not the free-fall time, but rather is the time taken for the
H$_{2}$ abundance to reach a maximum, which is simply the photodissociation time $t_{\rm dis}$. If 
$t_{\rm cool} \leq t_{\rm dis}$, then the gas will cool, but if $t_{\rm cool} \gg t_{\rm dis}$, it will remain warm.
As before, we know that 
\begin{equation}
t_{\rm cool} \propto n_{\rm H_{2}}^{-1},
\end{equation}
and it is also easy to show that the photodissociation time scales as
\begin{equation}
t_{\rm dis} \propto \frac{1}{G_{0}},
\end{equation}
independent of the value of the H$_{2}$ abundance. Therefore, the condition that $t_{\rm cool} = t_{\rm dis}$
implies that 
\begin{equation}
\frac{G_{0}}{n_{\rm H_{2}, eq}} = \mbox{constant}.
\end{equation}
When $G_{0} / n$ is large, this scales with
$n$ and $G_{0}$ as
\begin{equation}
n_{\rm H_{2}, eq} = \frac{n^{3}}{G_{0}^{2}},
\end{equation}
and hence our condition that $t_{\rm cool} = t_{\rm dis}$ is implies that 
\begin{equation}
\frac{G_{0}^{3}}{n^{3}} = \mbox{constant},
\end{equation}
and hence that 
\begin{equation}
\frac{G_{0}}{n} = \mbox{constant},
\end{equation}
in agreement with our simulation results.

\subsection{Varying the model parameters}
\label{res:ion}
\subsubsection{Initial ionization}
So far, we have assumed, for simplicity, that the gas is initially fully ionized. However, we know that in 
practice this is an overestimate, as the typical ionization fraction in the warm neutral medium is closer
to $x \sim 0.03$ \citep{w95}. It is therefore important to establish whether this simplification significantly affects our results. 
We have therefore run a series of models that include both metal and H$_{2}$ cooling, but that start
with fractional ionizations given by the following expression
\begin{equation}
x_{0} = \mbox{min} \left[1, 0.02 n^{-1/2} \left(\frac{\zeta_{\rm H}}{10^{-16} \: {\rm s^{-1}}} \right)^{1/2} \right],
\end{equation}
where $\zeta_{\rm H}$ is the cosmic ray ionization rate of atomic hydrogen. This expression yields values for
$x_{0}$ that are close to the equilibrium value that we would obtain if cosmic ray ionization and radiative
recombination were the only processes acting to change the ionization state of the gas. 

\begin{figure}
\includegraphics[width=0.4\textwidth]{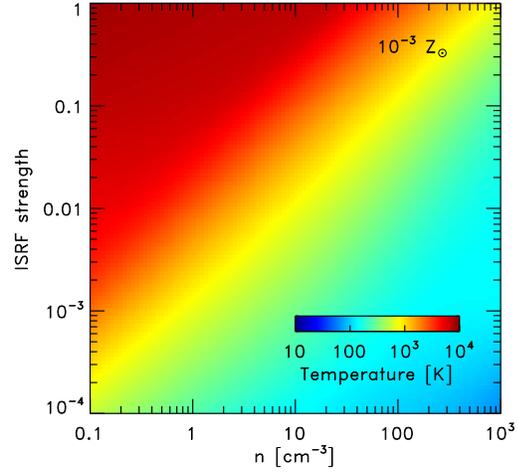}
\caption{As Figure~\ref{with-H2}, but for a set of simulations with
metallicity ${\rm Z} = 10^{-3} \: {\rm Z_{\odot}}$ that was run using a much 
lower value for the initial fractional ionization  of the gas.  \label{lowx}}
\end{figure}

We find that in general, the change in the initial ionization fraction has
only a minor effect on the ability of the gas to cool. As an example, we show in Figure~\ref{lowx} 
the results that we obtain for a run with ${\rm Z} = 10^{-3} \: {\rm Z_{\odot}}$ and a 
low initial fractional ionization. If we compare this with the corresponding panel in Figure~\ref{with-H2}, 
we see that there are only a few differences -- the region that is unable to cool has become slightly larger, 
and the final temperature reached by gas with high $n$ and low $G_{0}$ is also somewhat larger, 
but on the whole, the behaviour is very similar to that in the high ionization fraction case. 

In the regions of parameter space dominated by metal-line cooling, it is easy to understand why changing 
the initial fractional ionization has so little effect. In these simulations, the cooling rate is sensitive to the
fractional ionization only when the latter is very large \citep{dm72}. Therefore, although the cooling
rate in the high ionization simulations is initially larger than that in the low ionization simulations, the
cooling rates converge as the gas recombines and $x$ falls towards its equilibrium value. 

It is slightly more surprising that changing $x_{0}$ has such a limited effect in the case where H$_{2}$ 
cooling dominates, since as we have seen, the H$_{2}$ formation rate in these runs depends on the
fractional ionization. However, in practice, the dependence of the H$_{2}$ formation rate on $x$ 
becomes quite weak for $x > 0.03$, owing to the influence of the mutual neutralization reaction
\begin{equation}
{\rm H^{-}} + {\rm H^{+}} \rightarrow {\rm H} + {\rm H}.
\end{equation}
When $x$ is large, most of the H$^{-}$ ions formed by reaction~\ref{ra} are destroyed by this reaction
and do not survive for long enough to form H$_2$. Consequently, within this regime, 
further increasing $x$ has only a minor effect on the H$_{2}$ formation rate and the peak H$_{2}$
abundance. For this reason, the effect that decreasing $x_{0}$ has on the peak H$_{2}$ abundance
is much smaller than one might initially expect, thereby explaining why this change has only a limited
effect on the ability of the gas to cool.

\subsubsection{Dust-to-gas ratio}
\label{res:dust}
In most of our models, we assume that the dust-to-gas ratio, ${\cal D}$, scales linearly with 
metallicity. This appears to be a good assumption for galaxies with metallicities close to that of 
the Milky Way \citep{sand12}. However, there are observational indications
that at metallicities below around $0.3 \: {\rm Z_{\odot}}$, ${\cal D}$ falls off more rapidly with 
decreasing metallicity than predicted by a simple linear scaling \citep{gala11,hc12}. We have
therefore examined the effect of adopting a scaling for ${\cal D}$ that better matches the 
observational data, namely
\begin{equation}
{\cal D} = \left \{ \begin{array}{lr}
{\cal D_{\odot}} \left(\frac{{\rm Z}}{{\rm Z_{\odot}}} \right) & {\rm Z} \geq 0.3 \, {\rm Z_{\odot}} \\
0.3 \, {\cal D_{\odot}} \left(\frac{{\rm Z}}{{\rm 0.3 \: Z_{\odot}}} \right)^{2} & {\rm Z} < 0.3 \, {\rm Z_{\odot}}
\end{array}
\right.  \label{dust_scale}
\end{equation}

\begin{figure*}
\includegraphics[width=0.33\textwidth]{figures/final_temp-Z-1.eps}
\includegraphics[width=0.33\textwidth]{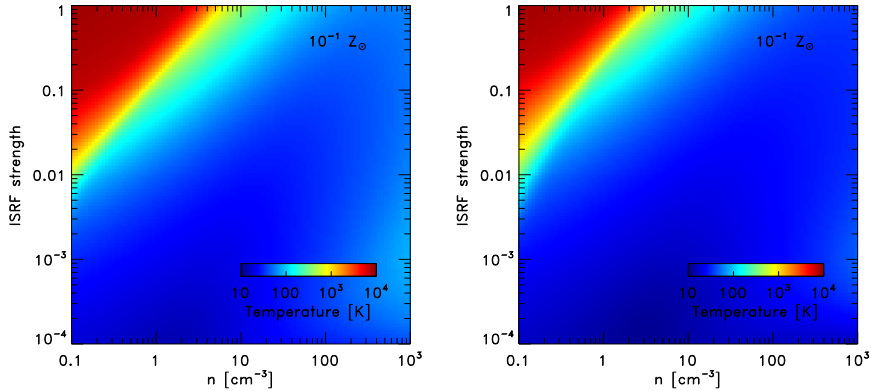}
\caption{Gas temperature at $t = t_{\rm ff}$, computed as a function of the
number density of hydrogen nuclei, $n$, and the strength of the interstellar 
radiation field in units of the standard value, $G_{0}$, for two runs
with metallicities ${\rm Z} = 0.1 \: {\rm Z_{\odot}}$. The left-hand panel
shows the results from a run using a linear scaling of the dust-to-gas
ratio, ${\cal D}$, with metallicity, while the right-hand panel shows the
results from a run using the steeper scaling given by Equation~\ref{dust_scale}. 
Cooling is slightly more effective in the run with less dust, but the difference
is relatively small.
\label{new-dust}}
\end{figure*}

We find that changing the dust scaling in this fashion has little effect on our results. Obviously,
for metallicities ${\rm Z} \geq 0.3 \: {\rm Z_{\odot}}$, there is no change in ${\cal D}$ from the
values used with our default set-up, and hence no change in behaviour. If we drop the 
metallicity to ${\rm Z} = 0.1 \: {\rm Z_{\odot}}$,  we do see a slight change in behaviour, as
illustrated in Figure~\ref{new-dust}: the gas in the simulation with less dust cools more readily 
than the gas in the simulation with more dust. This is a consequence of the fact that photoelectric
heating is still an important heat source in the diffuse ISM at these metallicities. By reducing
the dust abundance, we reduce the photoelectric heating rate, making it easier for the gas
to cool. Nevertheless, the difference between the two runs is small, and grows smaller still
if we further reduce the metallicity. 
We can therefore be confident that
the main results of this study are robust against uncertainties in the nature of the dust in 
low metallicity galaxies.

\subsubsection{H$_{2}$ self-shielding}
\label{res:h2ss}
\begin{figure*}
\includegraphics[width=0.33\textwidth]{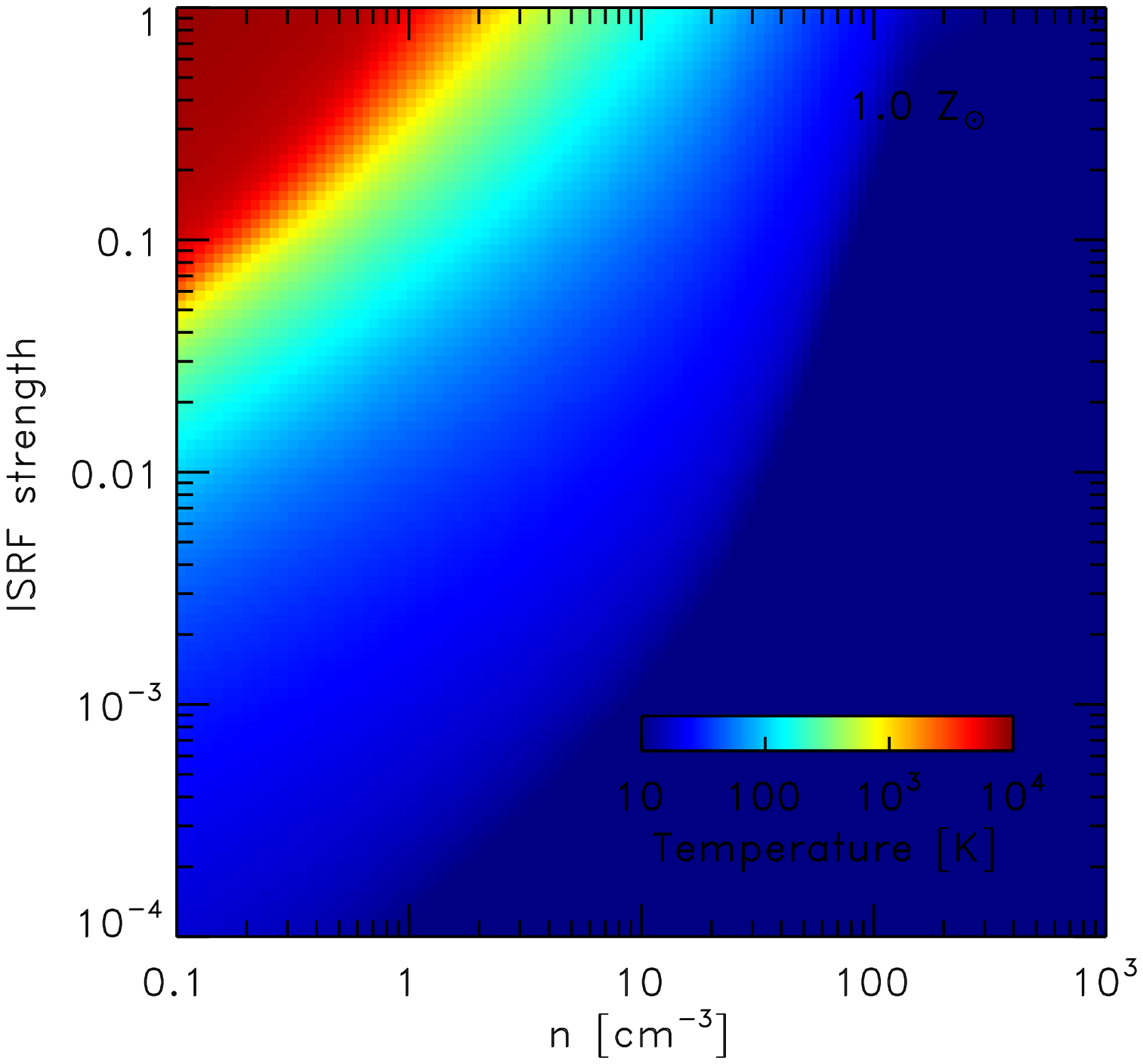}   
\includegraphics[width=0.33\textwidth]{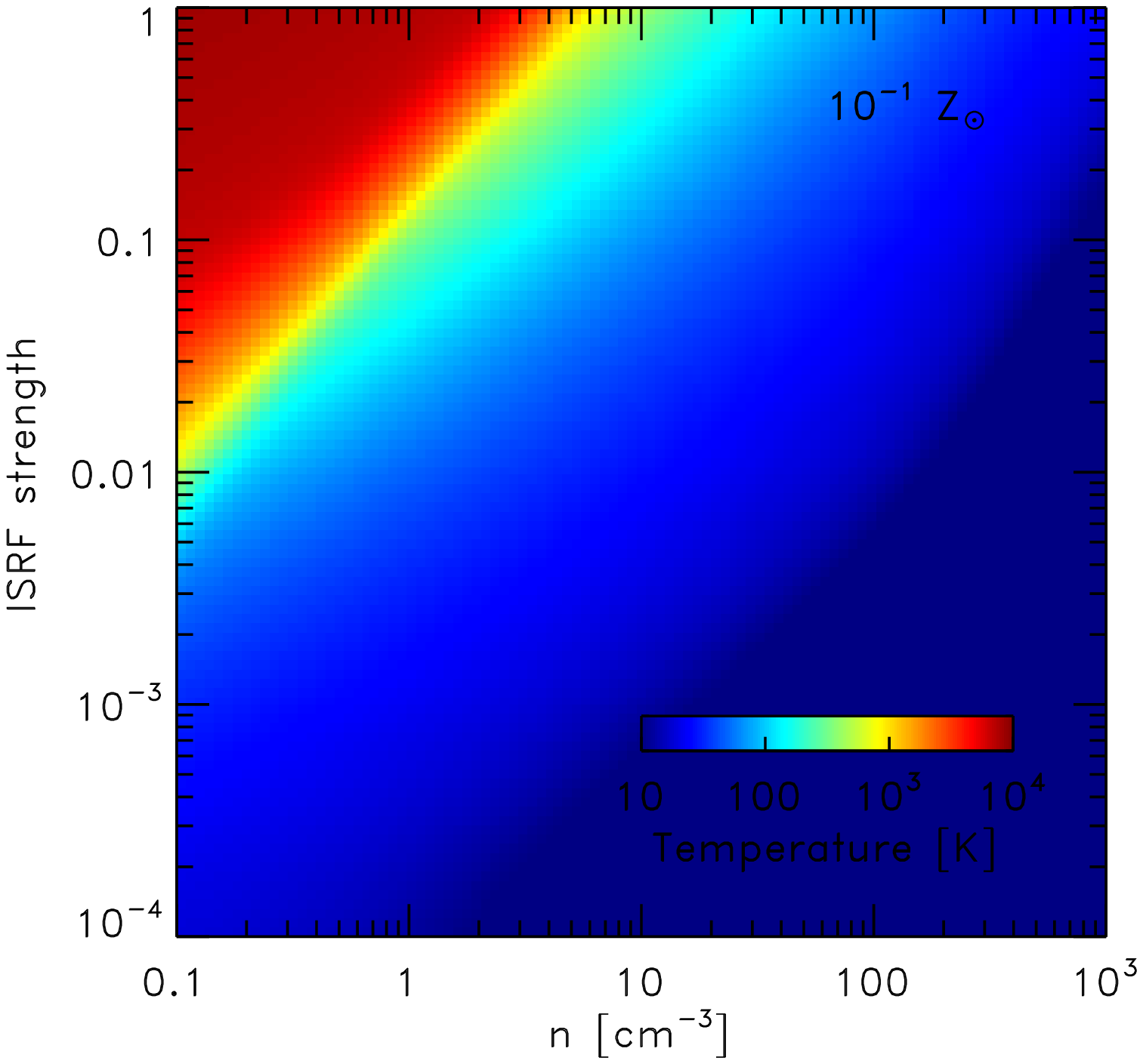}
\includegraphics[width=0.33\textwidth]{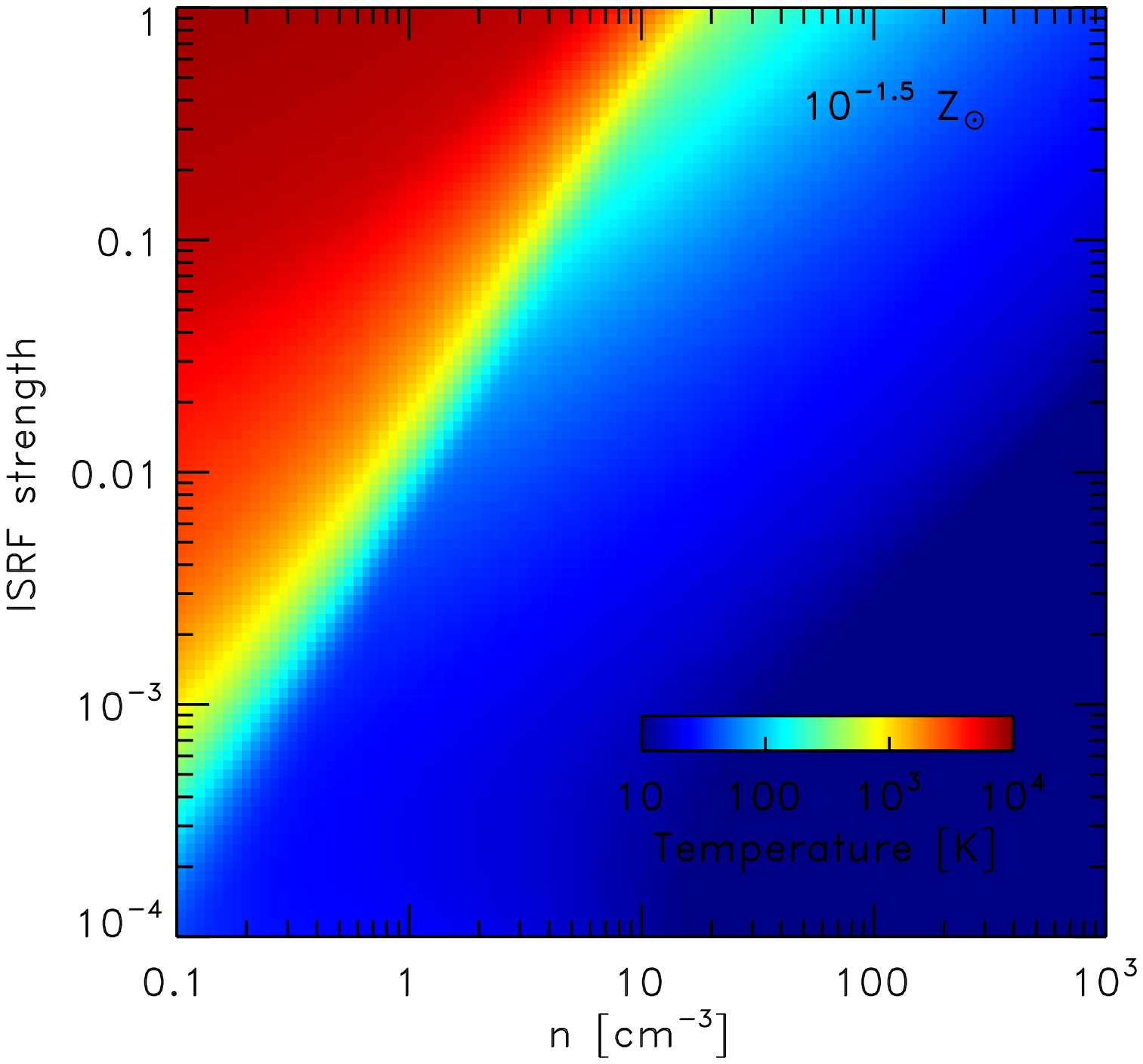}
\includegraphics[width=0.33\textwidth]{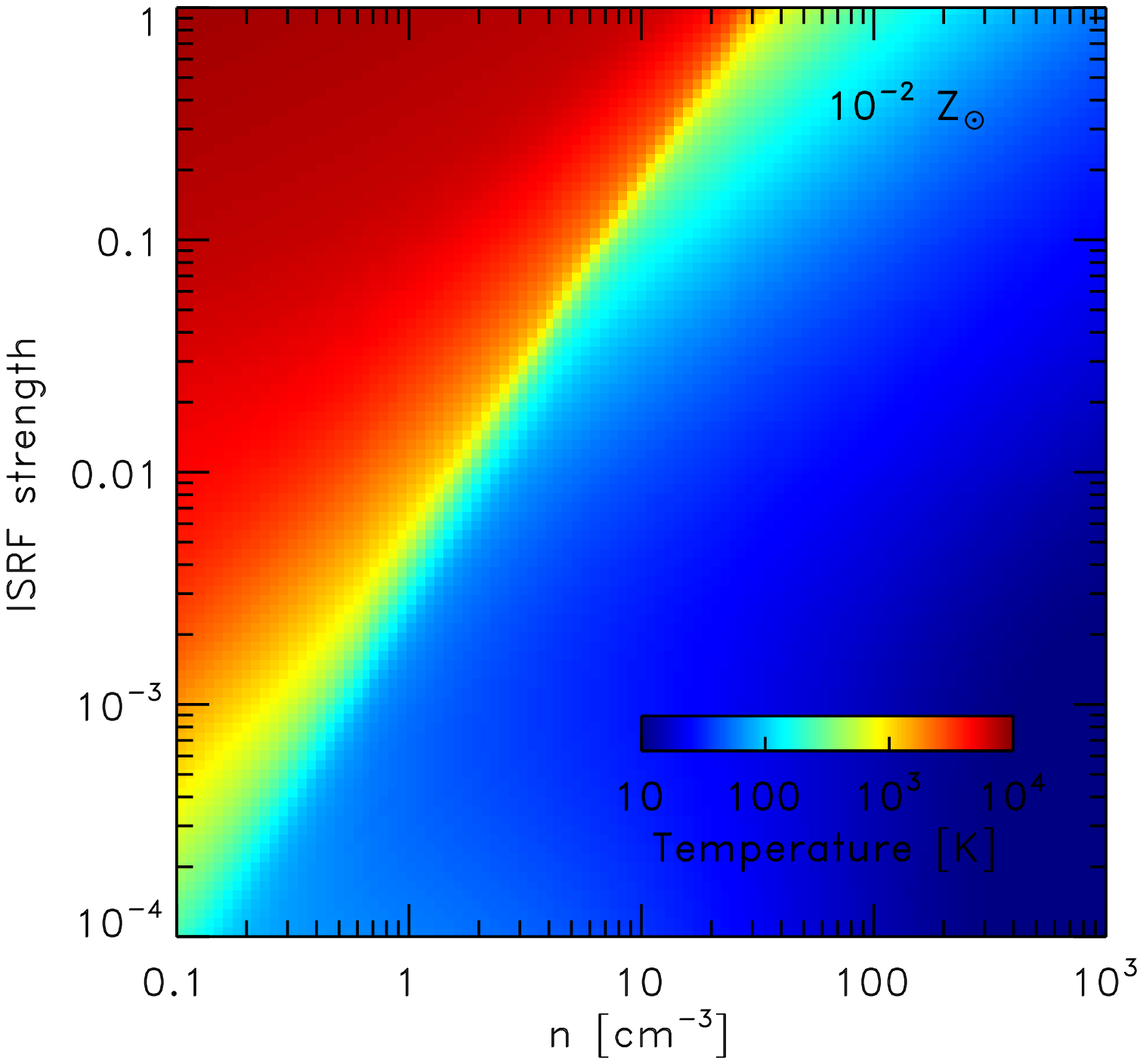}
\includegraphics[width=0.33\textwidth]{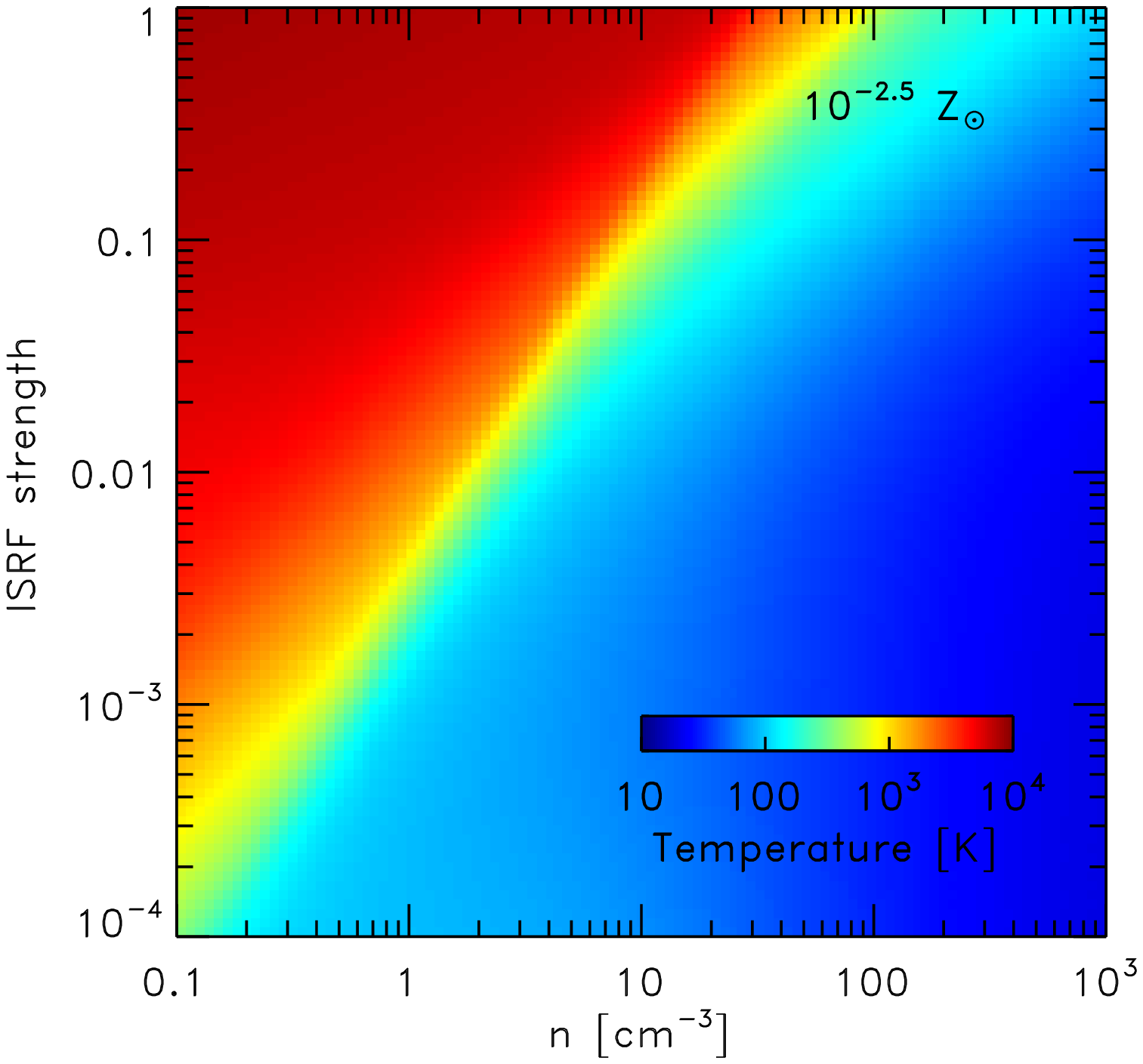}
\includegraphics[width=0.33\textwidth]{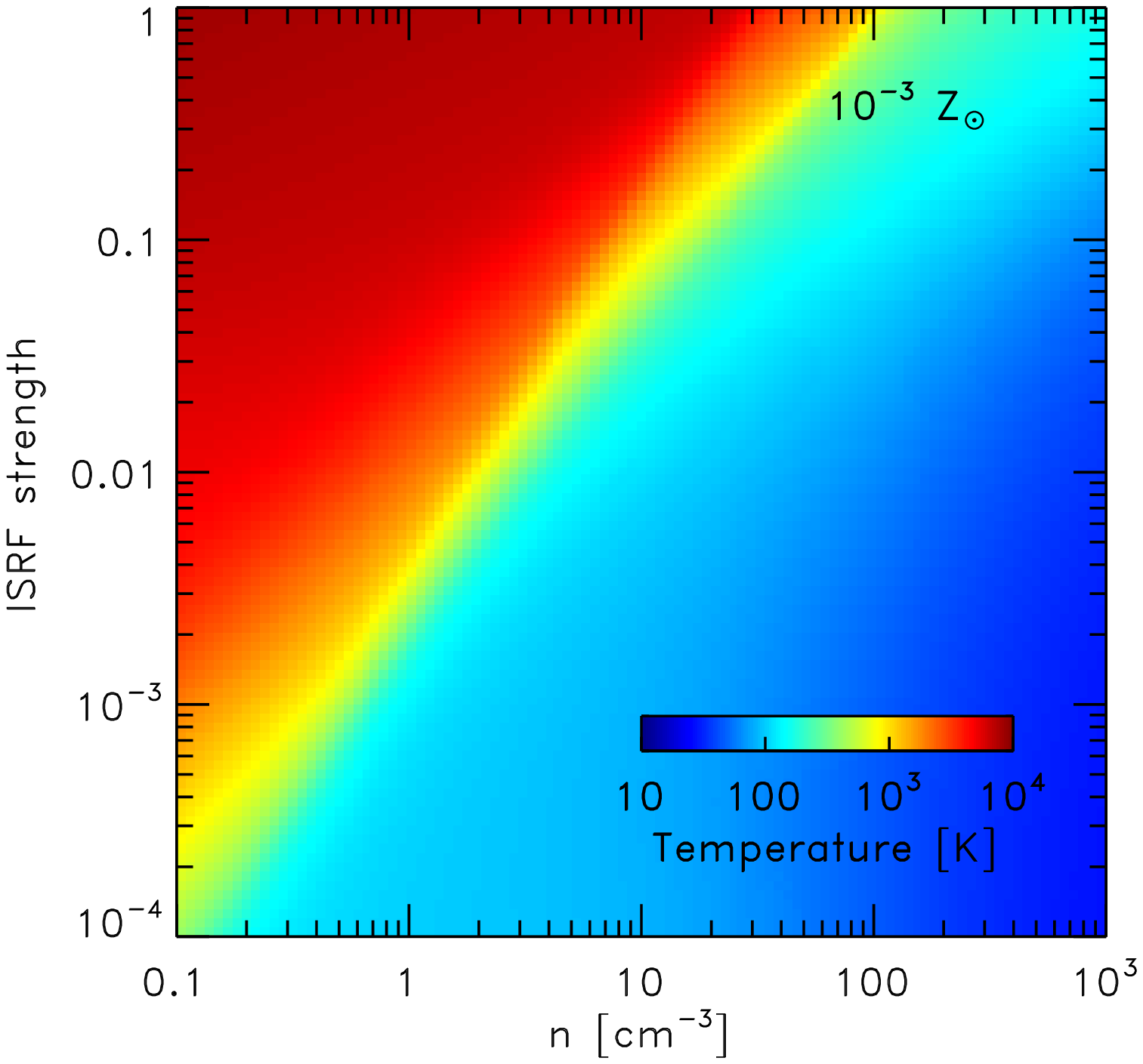}
\includegraphics[width=0.33\textwidth]{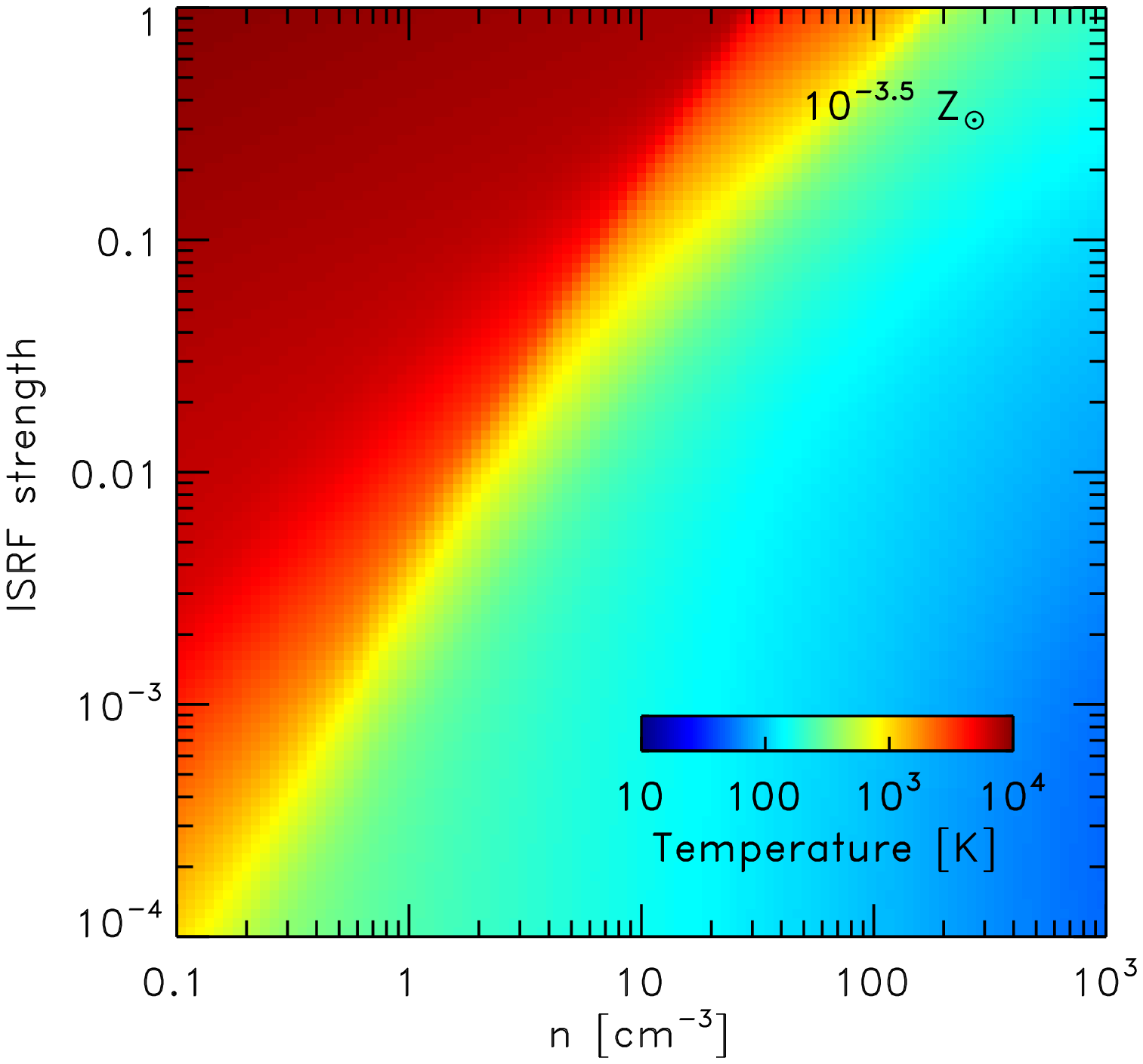}
\includegraphics[width=0.33\textwidth]{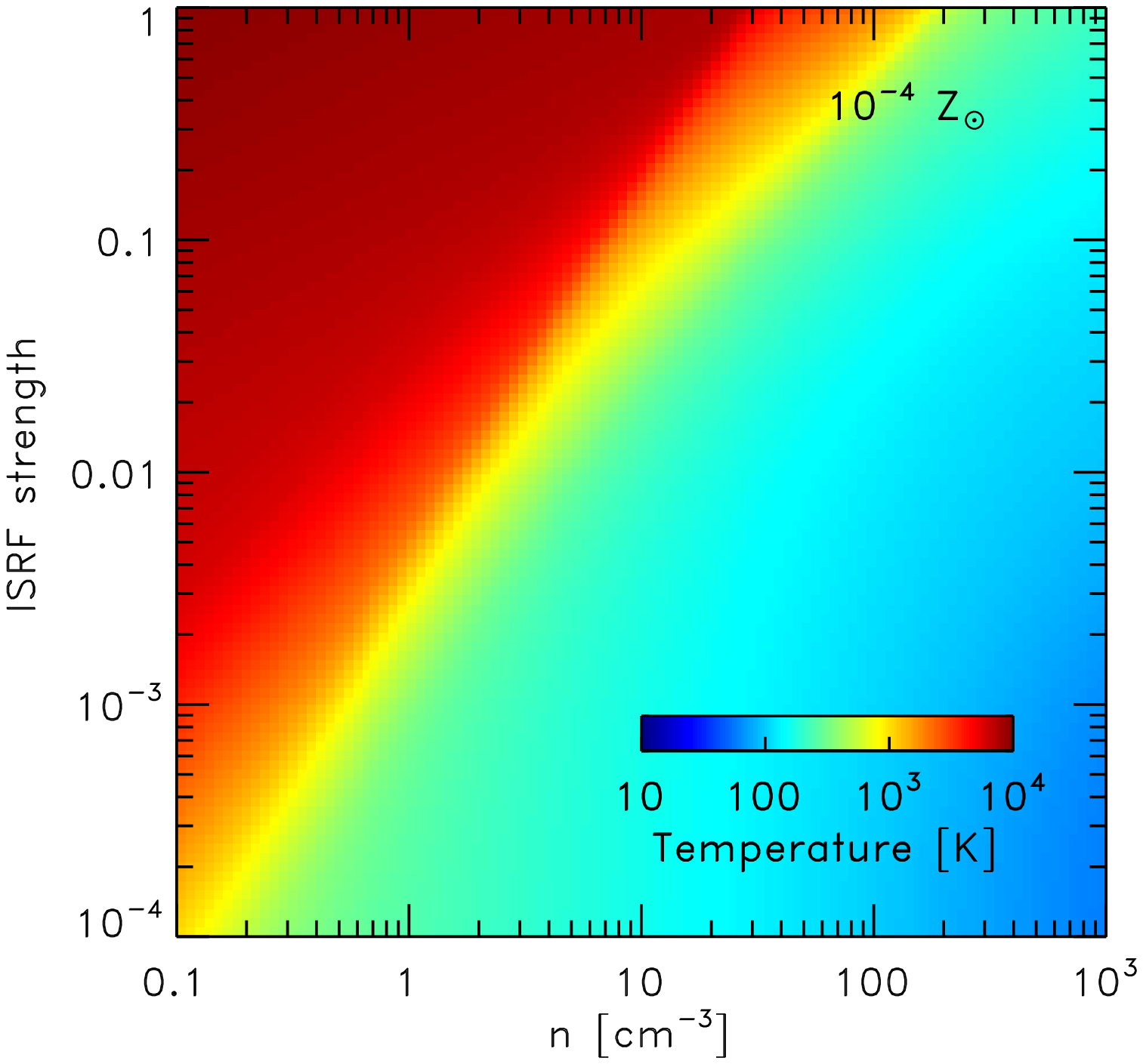}
\caption{As Figure~\ref{with-H2}, but for a set of simulations that include the
effects of H$_{2}$ self-shielding, with $L_{\rm ss} = 10 \: {\rm pc}$. \label{H2-shield}}
\end{figure*}

Up to this point, we have neglected the effects of H$_2$ self-shielding. However, it is not
immediately obvious that this is a good approximation. The H$_{2}$ fractions that we find
in the diffuse gas are small, but even H$_{2}$-poor gas can produce a significant amount
of self-shielding if its total column density is large enough. In the real ISM, the amount of
self-shielding that we have at any given point depends on the distribution of H$_{2}$
column densities around that point, as well as the velocity field of the gas. Our simple
one-zone model does not allow us to accurately account for these effects, and so to
investigate the influence of self-shielding we instead adopt the approximation discussed
in Section~\ref{sec:num}: we assume that the H$_{2}$ column density is directly related to 
the H$_{2}$ number density via
\begin{equation}
N_{\rm H_{2}} = n_{\rm H_{2}} L_{\rm ss}.   \label{lss}
\end{equation}
In Figure~\ref{H2-shield}, we show the results we obtain from a set of runs performed 
using this approximation, with $L_{\rm ss} = 10 \: {\rm pc}$. 

Comparing the results plotted in Figures~\ref{with-H2} and \ref{H2-shield}, we see
that at metallicities ${\rm Z} \geq 10^{-1.5} \: {\rm Z_{\odot}}$, the inclusion of H$_{2}$ 
self-shielding has little effect on the final gas temperature. In these simulations,
the regions in $G_{0}$--density parameter space where cooling is inefficient 
also correspond to regions where the peak H$_{2}$ fraction is very small. The
H$_{2}$ column density in these regions therefore never becomes large enough 
to produce significant self-shielding of H$_{2}$, and the final result is therefore
the same in simulations with and without this effect included. At higher densities
and lower values of G$_{0}$, the peak H$_2$ fraction is much higher and the gas
does become able to self-shield, but this occurs in regions that can already cool
efficiently, and so once again the inclusion of H$_{2}$  self-shielding has little 
effect on the outcome.  The only significant difference between the two sets of runs
is in the minimum temperature reached by the gas with high density and low G$_{0}$:
the inclusion of H$_{2}$ self-shielding allows this gas to cool to slightly lower 
temperatures than before. However, we caution that this result should be treated with
caution as it is likely that in reality this strongly cooling gas will change its density
in response to the loss of thermal support, rather than remaining at constant density as
assumed in our model.

In the lower metallicity runs, ${\rm Z} \leq 10^{-2} \: {\rm Z_{\odot}}$, the inclusion
of H$_{2}$ self-shielding has a more pronounced effect. The boundary in 
G$_{0}$--density space between regions that can and cannot cool becomes 
much sharper and also slightly steeper, corresponding to a line where
$G_{0} / n^{4/3} \sim {\rm constant}$, rather than to a line of constant $G_{0} / n$
as in the runs without self-shielding. The fact that the transition becomes sharper
is easy to understand: it is simply a consequence of the sudden jump in the 
equilibrium abundance of H$_{2}$ that occurs once the gas becomes able
to self-shield. This jump occurs because the onset of self-shielding leads to
a drop in the photodissociation rate, which leads to an increase in the equilibrium
H$_{2}$ abundance, which leads to an increase in the H$_{2}$ column density
and hence a further drop in the photodissociation rate, etc. This positive feedback
drives the equilibrium H$_{2}$ abundance sharply upwards once we enter the
self-shielding regime. 

The steepening of the transition is a consequence of the fact that by including
self-shielding, we are including another source of density dependence: the
amount by which self-shielding reduces the H$_{2}$ photodissociation rate
depends on N$_{\rm H_{2}}$, which in turn depends on the density via
Equation~\ref{lss}. Therefore, in gas with any given value of $G_{0} / n$, 
self-shielding, and hence H$_{2}$ cooling, will always be more effective
at large $n$ (and hence large $G_{0}$) than at small $n$. 

We have also examined the effect of varying $L_{\rm ss}$. We find, as
one might expect, that decreasing $L_{\rm ss}$ leads to an increase in the 
size of the region that does not cool, whereas increasing $L_{\rm ss}$ has
the opposite effect. However, in both cases the effect is fairly small. The
reason for this is that $n_{\rm H_{2}}$ varies strongly as a function of $G_{0}$
and $n$, as we have already seen, and hence the value of $N_{\rm H_{2}}$ 
at a given point in $G_{0}$--density space depends far more on the value
of $n_{\rm H_{2}}$ than on the value of $L_{\rm ss}$. Our main results are
therefore robust against changes in $L_{\rm ss}$.
 
\section{Discussion}
At first sight, it might appear that our results are in conflict with previous studies
of star formation in metal-poor systems that typically find that metal-line cooling
dominates over H$_{2}$ cooling at metallicities above a critical value of around
$10^{-3.5} \: {\rm Z_{\odot}}$ \citep[see e.g.][]{bfcl01,bl03,fjb07,smith09}, and
that dust cooling can dominate at even lower metallicities \citep{sch02,om05,clark08}. 
However, these studies were primarily concerned with higher 
density gas than is treated in this paper. For example, \citet{fjb07} derive their
constraints on the metallicity required to provide effective fine-structure cooling
by considering the behaviour of gas at a density of $10^{4} \: {\rm cm^{-3}}$ 
and a temperature of 200~K without addressing how the gas gets to this density 
and temperature in the first place. Similarly, studies of dust cooling in very 
metal-poor systems find that dust dominates the cooling only at very high densities 
\citep{dopcke11,dopcke13}.

In order to reach the densities at which metal cooling begins to dominate
in these metal-poor systems, which are orders of magnitude higher than typical ISM 
densities, some other form of cooling is required. One possibility is Lyman-$\alpha$ 
cooling from atomic hydrogen, but this maintains the gas temperature at close to 
$10^{4} \: {\rm K}$, resulting in a very large Jeans mass ($M_{\rm J} \sim 2 \times 10^{7} n^{-1/2} 
\: {\rm M_{\odot}}$). 
This will therefore be an important process only in relatively massive systems,
such as protogalaxies illuminated by strong UV radiation fields 
\citep[see e.g.][]{osh08}. What our results demonstrate is that when the ambient 
radiation field is relatively weak, molecular hydrogen alone can cool the gas in less 
than a free-fall time, thereby enabling it to collapse to the higher densities required 
for efficient metal cooling.

It is also interesting to compare our results for solar metallicity gas with those
presented by \citet{gk11}. In their simulations, \citet{gk11} find that the
H$_2$ cooling rate is almost an order of magnitude larger than the metal-line
cooling rate at temperatures of around 1000~K, and hence conclude that 
H$_2$ dominates the thermal evolution of the warm gas. On the other hand,
in our solar metallicity simulations, we find that H$_{2}$ never provides more
than about 10\% of the total cooling, even in runs in which we account for 
the effects of H$_{2}$ self-shielding. The reason for this discrepancy is not
entirely clear. Our treatment of H$_{2}$ cooling differs from that in \citet{gk11}
owing to the changes we have made to our treatment of H$_{2}$-proton and 
H$_{2}$-electron collisions (see Section~\ref{sec:num}), but the overall change is at
most a factor of three, far too little to account for the difference in our results. 
\citeauthor{gk11} also make use of a different treatment of metal-line cooling, 
based on \citet{pen70} and \citet{dm72}, but do not provide enough information 
on the ionization state of the gas in their models to allow us to judge how 
significantly their total metal-line cooling rate differs from ours. 

Finally, it is possible that the initial conditions that we adopt for the warm gas in
our present study do not properly reflect the conditions that \citet{gk11} find in
their 3D hydrodynamical simulation. Specifically, we assume that the gas is
initially completely devoid of H$_{2}$. This means that the amount of H$_2$
available to participate in the cooling cannot be larger than the amount that 
can form within a cooling time, even if the equilibrium H$_{2}$ abundance is
in principle much larger. In the \citet{gk11} simulations, some of their warm gas
will have previously been cold, dense gas, and it may therefore retain some of
the H$_{2}$ that it formed during this period. If so, then this would naturally lead
to higher H$_{2}$ cooling rates than we find in our simulations. 

\section{Summary}
\label{sec:sum}
Our results demonstrate that H$_{2}$ cooling can potentially play an important role 
in regulating the temperature of the diffuse ISM, but only if two important conditions are met. 
First, the metallicity must be below $0.1 \: {\rm Z_{\odot}}$, as at higher metallicities, 
metal line cooling dominates throughout the parameter space considered here,
rendering the presence or absence of H$_{2}$ irrelevant as far as the thermal evolution
of the gas is concerned. Second, the ratio of the ISRF strength to the gas density must also 
be low, typically $G_{0} / n \sim 0.01$, or else not enough H$_{2}$ will survive to provide 
effective cooling. 

In the extreme case where $G_{0} = 0$, i.e.\ when there is no interstellar
radiation field present, we predict that H$_{2}$ will dominate over a wide range
of physical conditions, in good agreement with the results of the previous study 
of \citet{jappsen07}.  On the other hand, when the radiation field is strong, our results 
show that H$_{2}$ cooling will be ineffective unless the gas density is already large.
In a highly turbulent system \citep[e.g.][]{walch11} or in one which is undergoing large-scale 
gravitational collapse, such as a high-redshift protogalaxy \citep[e.g.][]{bfcl01,smith09}, 
it is possible that the required densities could be achieved without the need for efficient cooling 
at $T < 10^{4} \: {\rm K}$, but in more quiescent systems, it is likely that the required value of
$G_{0} / n$ can be achieved only when $G_{0}$ itself is small, i.e.\ when the
interstellar radiation field is weak.

Our results therefore suggest that although H$_{2}$ cooling may enable star 
formation to occur in low metallicity systems that otherwise would be unable
to cool, the effect will be strongly self-limiting: if too many stars form, the interstellar
radiation field will become too strong, inhibiting further cooling and star formation.

\section*{Acknowledgments}
The authors would like to thank F.~Bigiel, E.~Ostriker and S.~Walch for useful discussions on
the topic of cooling in the ISM. They would also like to thank the anonymous referee for a useful
and constructive report. Financial support for this work was provided by the Deutsche 
Forschungsgemeinschaft (DFG) via SFB 881``The Milky Way System'' (sub-projects B1 and B2), 
and from the Baden-W\"{u}rttemberg-Stiftung by contract research via the programme 
Internationale Spitzenforschung II (grant P- LS-SPII/18). PCC also acknowledges support 
from grant CL 463/2-1, part of the DFG priority program 1573 ``Physics of the Interstellar Medium''.


\begin{thebibliography}{}

\bibitem[Abrahamsson, Krems \& Dalgarno(2007)]{akd07}
Abrahamsson, E., Krems, R.~V., \& Dalgarno, A.\ 2007, ApJ, 654, 1171

\bibitem[Aykutalp \& Spaans(2011)]{as11}
Aykutalp, A., \& Spaans, M.\ 2011, ApJ, 737, 63

\bibitem[Bigiel et al.(2008)]{bigiel08}
Bigiel, F., Leroy, A., Walter, F., Brinks, E., de Blok, W.~J.~G., Madore, B.,
\& Thornley, M.~D.\ 2008, AJ, 136, 2846

\bibitem[Bigiel et al.(2011)]{bigiel11}
Bigiel, F., et~al.\ 2011, ApJ, 730, L13

\bibitem[Bromm et~al.(2001)]{bfcl01}
Bromm, V., Ferrara, A., Coppi, P.~S., \& Larson, R.~B., 2001, MNRAS, 328, 969

\bibitem[Bromm \& Loeb(2003)]{bl03}
Bromm, V., \& Loeb, A.\ 2003, Nature, 425, 812

\bibitem[Cazaux \& Spaans(2004)]{cs04}
Cazaux, S., \& Spaans, M.\ 2004, ApJ, 611, 40

\bibitem[Clark, Glover \& Klessen(2008)]{clark08}
Clark, P.~C., Glover, S.~C.~O., \& Klessen, R.~S.\ 2008, ApJ, 672, 757

\bibitem[Clark et~al.(2011)]{clark11}
Clark, P.~C., Glover, S.~C.~O., Klessen, R.~S., \& Bromm, V.\ 2011, ApJ, 727, 110

\bibitem[Croft, Dickinson \& Gadea(1999)]{cdg99}
Croft, H., Dickinson, A.~S., \& Gadea, F.~X.\ 1999, MNRAS, 304, 327

\bibitem[Dalgarno \& McCray(1972)]{dm72}
Dalgarno, A., McCray, R.~A.\ 1972, ARA\&A, 10, 375

\bibitem[Draine(1978)]{dr78}
Draine, B.~T. 1978, ApJS,  36, 595

\bibitem[Draine \& Bertoldi(1996)]{db96}
Draine, B.~T., \& Bertoldi, F.\ 1996, ApJ, 468, 269

\bibitem[Dopcke et~al.(2011)]{dopcke11}
Dopcke, G., Glover, S.~C.~O., Clark, P.~C., \& Klessen, R.~S.\ 2011, ApJ, 729, L3

\bibitem[Dopcke et~al.(2013)]{dopcke13}
Dopcke, G., Glover, S.~C.~O., Clark, P.~C., \& Klessen, R.~S.\ 2013, ApJ, 766, 103


\bibitem[Frebel, Johnson \& Bromm(2007)]{fjb07}
Frebel, A., Johnson, J.~L., \& Bromm, V.\ 2007, MNRAS, 380, 40

\bibitem[Galametz et~al.(2011)]{gala11}
Galametz, M., Madden, S.~C., Galliano, F., Hony, S., Bendo, G.~J., \& Sauvage, M.\
2011, A\&A, 532, 56 

\bibitem[Gerlich(1990)]{ger90}
Gerlich, D., 1990, J.\ Chem.\ Phys., 92, 2377

\bibitem[Glover(2003)]{glo03}
Glover, S.~C.~O., 2003, ApJ, 584, 331

\bibitem[Glover \& Abel(2008)]{ga08}
Glover, S.~C.~O., \& Abel, T., 2008, MNRAS, 388, 1627

\bibitem[Glover \& Clark(2012a)]{gc12a}
Glover, S.~C.~O., \& Clark, P.~C.\ 2012a, MNRAS, 421, 9

\bibitem[Glover \& Clark(2012b)]{gc12b}
Glover, S.~C.~O., \& Clark, P.~C.\ 2012b, MNRAS, 426, 377

\bibitem[Glover et~al.(2010)]{glo10}
Glover, S.~C.~O., Federrath, C., {Mac Low}, M.-M., \& Klessen, R.~S.\ 2010, MNRAS, 404, 2

\bibitem[Glover \& Jappsen(2007)]{gj07}
Glover, S.~C.~O., \& Jappsen, A.-K.\ 2007, ApJ, 666, 1

\bibitem[Gnedin \& Kravtsov(2011)]{gk11}
Gnedin, N.~Y., \& Kravtsov, A.~V., 2011,  ApJ, 728, 88

\bibitem[Habing(1968)]{habing68}
Habing, H.~J. 1968, Bull. Astron. Inst. Netherlands,  19, 421

\bibitem[Herrera-Camus et~al.(2012)]{hc12}
Herrera-Camus, R., et~al., 2012, ApJ, 752, 112

\bibitem[Hollenbach \& McKee(1979)]{hm79}
Hollenbach, D., McKee, C.~F.\ 1979, ApJS, 41, 555

\bibitem[Hollenbach \& McKee(1989)]{hm89}
Hollenbach, D., McKee, C.~F.\ 1989, 342, 306

\bibitem[Honvault et~al.(2011)]{hon11}
Honvault, P., Jorfi, M., Gonz\'{a}lez-Lezana, T., Faure, A., \& Pagani, L.\ 2011, Phys.\ Rev.\ Lett., 107, 023201

\bibitem[Honvault et~al.(2012)]{hon12}
Honvault, P., Jorfi, M., Gonz\'{a}lez-Lezana, T., Faure, A., \& Pagani, L.\ 2012, Phys.\ Rev.\ Lett., 108, 109903

\bibitem[Jappsen et~al.(2007)]{jappsen07}
Jappsen, A.-K., Glover, S.~C.~O., Klessen, R.~S., \& {Mac Low}, M.-M., 2007, ApJ, 660, 1332

\bibitem[Kreckel et~al.(2010)]{kreck10}
Kreckel, H., Bruhns, H., \v{C}\'{i}\v{z}ek, M., Glover, S.~C.~O., Miller, K.~A., Urbain, X., \& Savin, D.~W.\ 2010, Science, 329, 69

\bibitem[Krsti\'c(2002)]{kr02}
Krsti\'c, P.~S. 2002, Phys.\ Rev.\ A, 66, 042717

\bibitem[Krumholz, Leroy \& McKee(2011)]{klm11}
Krumholz, M.~R., Leroy, A.~K., \& McKee, C.~F.\ 2011, ApJ, 731, 25

\bibitem[Krumholz(2012)]{krum12}
Krumholz, M.~R., 2012, ApJ, 759, 9

\bibitem[Leroy et al.(2008)]{leroy08}
Leroy, A.~K., Walter, F.,  Brinks, E., Bigiel, F., de Blok, W.~J.~G., Madore, B.,
\& Thornley, M.~D.\ 2008, AJ, 136, 2782

\bibitem[Mathis, Mezger \& Panagia(1983)]{mmp83}
Mathis, J.~S., Mezger, P.~G., \& Panagia, N.\ 1983, A\&A, 128, 212

\bibitem[Oh \& Haiman(2002)]{oh02}
Oh, S.~P., \& Haiman, Z., 2002, ApJ, 569, 558

\bibitem[Omukai et~al.(2005)]{om05}
Omukai, K., Tsuribe, T., Schneider, R., \& Ferrara, A.\ 2005, ApJ, 626, 627

\bibitem[Omukai, Schneider \& Haiman(2008)]{osh08}
Omukai, K., Schneider, R., \& Haiman, Z.\ 2008, ApJ, 686, 801

\bibitem[Penston(1970)]{pen70}
Penston, M.~V.\ 1970, ApJ, 162, 771

\bibitem[Rees \& Ostriker(1977)]{ro77}
Rees, M.~J., \& Ostriker, J.~P.\ 1977, MNRAS, 179, 541

\bibitem[Sandstrom et~al.(2013)]{sand12}
Sandstrom, K.~M., et~al., 2013, ApJ, in press; arXiv:1212.1208

\bibitem[Schneider et~al.(2002)]{sch02}
Schneider, R., Ferrara, A., Natarajan, P., \& Omukai, K.\ 2002, ApJ, 571, 30

\bibitem[Schruba et~al.(2011)]{schruba11}
Schruba, A., et~al., 2011, AJ, 142, 37

\bibitem[Sembach et~al.(2000)]{sem00}
Sembach, K.~R., Howk, J.~C., Ryans, R.~S.~I., \& Keenan, F.~P.,
2000, ApJ,  528, 310

\bibitem[Shetty, Kelly \& Bigiel(2013)]{shetty13}
Shetty, R., Kelly, B.~C., \& Bigiel, F.\ 2013, MNRAS, 430, 288

\bibitem[Stenrup, Larson \& Elander(2009)]{sten09}
Stenrup, M., Larson, \AA, Elander, N.\ 2009, Phys.\ Rev.\ A, 79, A2713 

\bibitem[Smith et~al.(2009)]{smith09}
Smith, B.~D., Turk, M.~J., Sigurdsson, S., O'Shea, B.~W., \& Norman, M.~L.,
2009, ApJ, 691, 441

\bibitem[Tegmark et~al.(1997)]{teg97}
Tegmark, M., Silk, J., Rees, M.~J., Blanchard, A., Abel, T., \& Palla, F., 1997, ApJ, 474, 1

\bibitem[Urbain et~al.(2012)]{urb12}
Urbain, X., Lecointre, J., Mezdari, F., Miller, K.~A., \& Savin, D.~W.\ 2012, J.\ Phys.\ Conf.\ Ser., 388, 092004

\bibitem[Walch et~al.(2011)]{walch11}
Walch, S., W\"unsch, R., Burkert, A., Glover, S., \& Whitworth, A., 2011, ApJ, 733, 47

\bibitem[Wolcott-Green, Haiman \& Bryan(2011)]{wghb11}
Wolcott-Green, J., Haiman, Z., \& Bryan, G., 2011, MNRAS, 418, 838

\bibitem[Wolfire et~al.(1995)]{w95}
Wolfire, M.~G.,  Hollenbach, D., McKee, C.~F.,Tielens, A.~G.~G.~M., Bakes, E.~L.~O.\ 1995, ApJ, 443, 152

\bibitem[Wolfire et~al.(2003)]{w03}
Wolfire, M.~G., McKee, C.~F., Hollenbach, D., Tielens, A.~G.~G.~M.\ 2003, ApJ, 587, 278

\bibitem[Yoon et~al.(2008)]{yoon08}
Yoon, J.-S., Song, M.-Y., Han, J.-M., Hwang, S.~H., Chang, W.-S., Lee, B., \& Itikawa, Y. 2008, J.\ Phys.\ Chem.\ Ref.\ Data,  37, 913

\end{thebibliography}
\end{document}